\documentclass{article}

\usepackage[preprint]{neurips_2026}

\usepackage[utf8]{inputenc} %
\usepackage[T1]{fontenc}    %
\usepackage{hyperref}       %
\usepackage{url}            %
\usepackage{booktabs}       %
\usepackage{amsfonts}       %
\usepackage{nicefrac}       %
\usepackage{microtype}      %
\usepackage{xcolor}         %

\title{Extending Myerson's Optimal Auctions to Correlated Bidders via Neural Network Interpolation}

\author{%
  Mingyu Guo\textsuperscript{1} \quad
  Jiayuan Liu\textsuperscript{2} \quad
  Vincent Conitzer\textsuperscript{2} \\
  \textsuperscript{1}Adelaide University, Adelaide, Australia\\
  \textsuperscript{2}Carnegie Mellon University, Pittsburgh, USA\\
  \texttt{mingyu.guo@adelaide.edu.au}\\
  \texttt{jiayuan4@andrew.cmu.edu, conitzer@cs.cmu.edu}
}

\usepackage{subcaption}
\usepackage{graphicx}
\usepackage{amsmath}
\usepackage{amsthm}
\usepackage{amsfonts}
\usepackage{wrapfig}

\begin{document}

\maketitle

\begin{abstract}
We aim to design revenue-maximizing single-item auctions that are deterministic, strategy-proof, and ex post individually rational --- the quintessential fundamental model for optimal mechanism design. While Myerson's seminal work solved this model for independent bidders, straightforward extensions to correlated settings often result in non-monotonic, and therefore non-strategy-proof allocations. Critically, optimal allocation design for correlated bidders is NP-hard to approximate, rendering theoretical optimality guarantees in this domain computationally intractable.

We establish a new standard of empirical rigor for this theoretically hard model by proposing an empirical pipeline. We synthesize neural interpolation of Myerson's greedy allocation based on marginal profits (the correlated variant of virtual valuation), formal neural verification to enforce exact strategy-proofness, and a constructive repair procedure.

Empirically, our method is highly effective, consistently achieves near-optimal revenue across a wide range of distributions, including synthetic, adversarial, and real-world (consumer, financial, and industrial) datasets.  Compared to existing manual and neural baselines, our approach shows substantial improvement, often reducing the revenue gap by an order of magnitude. Our study is also the first to introduce (unattainable) greedy revenue as a rigorous upper bound for empirical benchmarking, providing a definitive quantitative performance measure where true optima remain theoretically elusive.

We demonstrate the generality of our approach by extending it to multi-unit auctions with unit demand and integrating our verification techniques into RegretNet to achieve exact strategy-proofness.
\end{abstract}

\section{Introduction}
We address the design of revenue-maximizing single-item auctions -- the canonical model that serves as the foundation of mechanism design. In his landmark work~\cite{Myerson81:Optimal}, Myerson demonstrated that for independent bidders, an allocation rule is strategy-proof if and only if it is {\em monotone}. By greedily allocating based on {\em (ironed) virtual valuations}, one naturally satisfies monotonicity, yielding a deterministic and ex-post individually rational optimal auction.

However, this theoretical elegance is the primary bottleneck for practitioners: {\bf correlation is prevalent in real-life auctions.} When one naturally extends Myerson's technique to correlated settings by greedily allocating based on {\em marginal profits} (the correlated variant of virtual valuations), the resulting allocations are frequently non-monotone. Theoretical attempts to find an exact characterization for this setting have hit a mathematical wall; {\bf designing an optimal allocation for correlated bidders is proven NP-hard to approximate}~\cite{Papadimitriou11:On,Caragiannis16:Limitations}, rendering global optimality guarantees computationally intractable.

Recognizing this intractability, our work shifts the focus from fighting NP-hardness to empirical and computational rigor. We argue that for the correlated setting -- the most natural extension of Myerson's foundational optimal auction model -- the goal should be {\bf computationally designed mechanisms that are verifiably strategy-proof and near-optimal}. We propose a new approach by synthesizing a pipeline of {\bf neural interpolation} of Myerson's greedy allocation, {\bf formal strategy-proofness verification}, and {\bf repair}.
This framework effectively provides a computational resolution to this foundational model -- a cornerstone of mechanism design -- for practical implementation
in real-life auctions where theoretical optima are unavailable. Our results demonstrate that we can recover a perfectly strategy-proof and near-optimal mechanism from Myerson's greedy allocation even when the underlying target is mathematically ``broken'' (i.e., non-monotone).
We summarize our contributions as follows:

\begin{figure*}[h]
\centering
\begin{subfigure}[b]{0.24\textwidth}
    \centering
    \includegraphics[width=\textwidth]{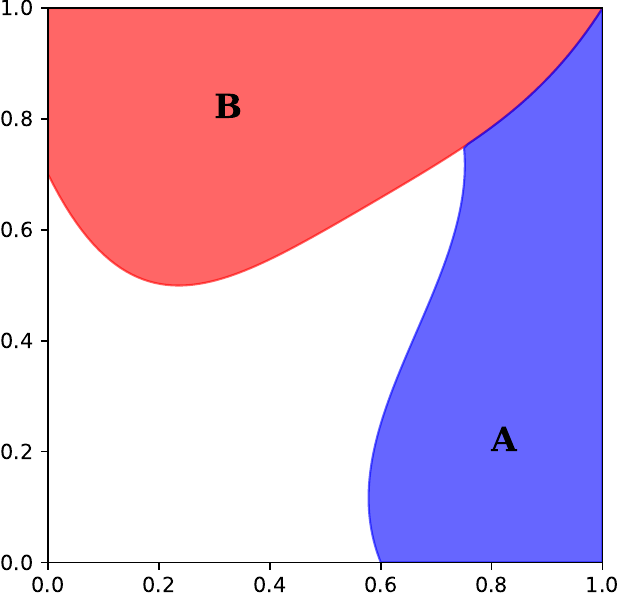}
    \caption{monotone allocation}
    \label{fig:mainfigure}
\end{subfigure}
\begin{subfigure}[b]{0.24\textwidth}
    \centering
    \includegraphics[width=\textwidth]{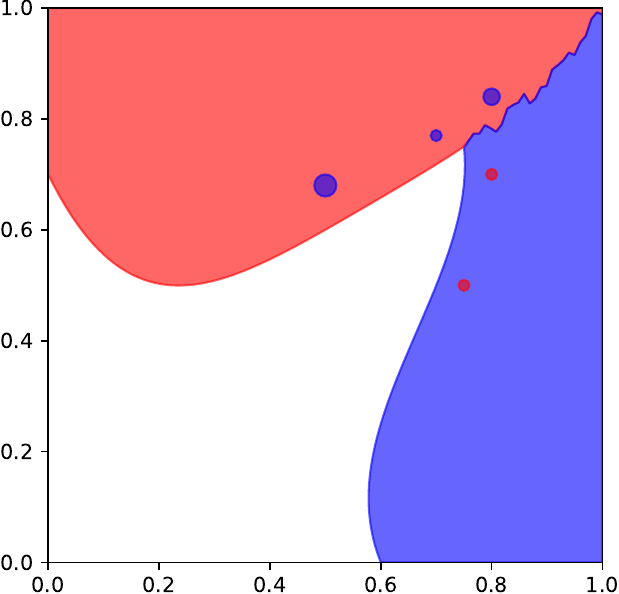}
    \caption{greedy allocation}
    \label{fig:greedy}
\end{subfigure}
\begin{subfigure}[b]{0.24\textwidth}
    \centering
    \includegraphics[width=\textwidth]{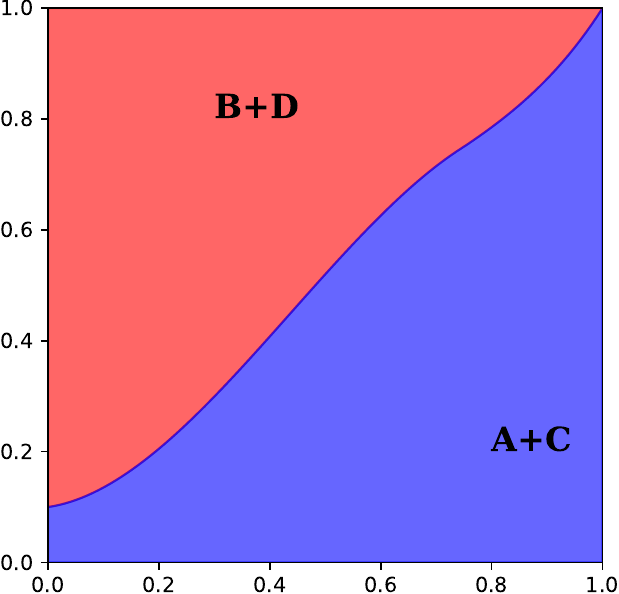}
    \caption{interpolation}
    \label{fig:learnt}
\end{subfigure}
\begin{subfigure}[b]{0.24\textwidth}
    \centering
    \includegraphics[width=\textwidth]{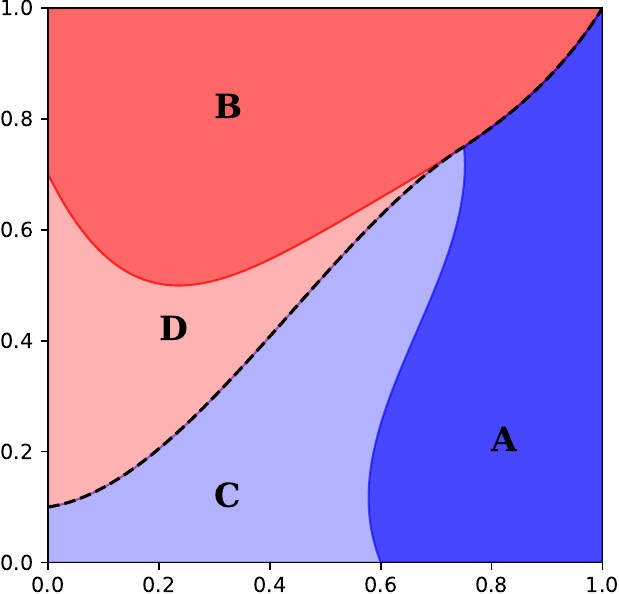}
    \caption{revenue fix}
    \label{fig:revenuefix}
\end{subfigure}
    \caption{
        Illustration of a 2-bidder setting where axes represent two bids in $[0,1]$. Blue and red regions denote wins for bidder 1 and 2, respectively. (a) monotone allocation: A strategy-proof allocation where a winner necessarily remains the winner after a unilateral bid increase. (b) greedy allocation (based on marginal profits): Non-monotone allocation that gives the revenue upper bound --- red and blue points enclosed in opposing regions represent monotonicity violations that preclude strategy-proofness. (c) (monotone) interpolation: A tiny neural network mimics the macro-trends of (b) while ``glossing over'' fine-grained violations due to its limited expressive capacity. This interpolation is formally verified to be monotone, thereby ensuring exact strategy-proofness.
    (d) revenue fix (leading to final allocation): The result after applying the revenue fix in Section~\ref{sec:vv}, which determines the exact shape of the ``no-sale'' (white) region to maximize expected revenue without altering the strategy-proofness established in the previous stage.}
\label{fig:storyillustration}
\end{figure*}

\vspace{.05in}
\noindent{\bf Contribution 1: Neural network interpolation as a verification tool for strategy-proofness.}

Given a specific correlated distribution, a natural first step is to check whether Myerson's greedy allocation remains monotone {\em despite the correlation}.  Note that it is infeasible to enumerate all bid profiles to verify monotonicity.
We train a neural network to interpolate Myerson's greedy allocation and then formally verify its monotonicity.\footnote{Although there are neural network architectures that ensure ``monotonicity'', such as the {\em min-max networks}~\cite{Sill97:Monotonic}, it is important to note that the notion of ``monotonicity'' represented by min-max networks is different from the concept of ``allocation monotonicity'' in mechanism design, despite the name similarity.  We defer the detailed discussion on this difference to Appendix~\ref{sec:altnet}.}

We use a multilayer perceptron (MLP) with ReLU activation to model the allocation function. This architecture allows us to {\em exactly verify} the monotonicity of a trained allocation using mixed-integer programming (MIP).  For training (i.e., the neural interpolation step), training samples are bid profiles drawn from the target distribution. We instruct the network to allocate to the highest marginal profit bidder.

Notably, we observe that for many correlated distributions, Myerson's greedy allocation is {\bf not} monotone, but the supervision result is a verifiably monotone interpolation. While the greedy allocation is rarely monotone, the deviations are often minimal.\footnote{For $2$-bidder cases where it is scalable to run the classical MIP-based automated mechanism design (AMD)~\cite{Conitzer02:Complexity}, we observe that greedy and AMD (strategy-proof therefore monotone) generally coincide, differing in only a small percentage of bid profiles, more in Appendix~\ref{subsec:ec}.}
The fact that a non-monotone target yields a monotone interpolation is a consequence of our network architecture as a {\bf ``detail-filtering'' regularizer}. Our MIP-based verification necessitates the use of relatively small networks. These ``tiny'' networks lack the expressive capacity to capture the fine-grained monotonicity violations of the greedy target; instead, they ``gloss over'' these violations to interpolate the underlying macro-trend. As illustrated in Figure~\ref{fig:storyillustration}, the interpolation (c) effectively filters the non-monotone noise of the greedy target (b) to recover a verifiably monotone allocation.

Finally, we propose that {\em neural network interpolation can serve as a general tool for verifying mechanism properties}, including but not limited to strategy-proofness. The mechanism may be expressed mathematically or in code.
Our pipeline utilizes neural interpolation as a bridge to formal verification, providing a robust framework to audit mechanism properties that are otherwise computationally inaccessible.

\vspace{.05in}
\noindent {\bf Contribution 2: A suite of monotonicity-seeking techniques, including counterexample-guided training and post-processing monotonicity fix, both enabled by the MLP+ReLU architecture.}

Besides verification, the MLP+ReLU architecture enables two monotonicity-seeking techniques.
First, when verification fails, the ``by-products'' of the verification MIP are bid profiles that violate monotonicity,
which can be used in {\em counterexample-guided training}~\cite{Sivaraman20:Counterexample-guided}.
The idea is to punish the violation in follow-up training until the violation disappears.
Second, if counterexample-guided training still fails to reach monotonicity, then we can implement
a post-processing {\em monotonicity fix} to ensure strategy-proofness.\footnote{Our monotonicity fix post-processes the allocation network to ensure strategy-proofness. GemNet~\cite{Wang24:GemNet}
proposed a similar approach that post-processes the payment network to ensure menu compatibility ({\em i.e.}, preventing two bidders from
winning the same item). Their method discretizes the bids
and then extend to general bids via Lipschitz smoothness. In contrast, our technique operates directly on continuous
values.}
We construct another MIP based on the network parameters, which decides how much to push up the winner's offer while fixing the other bids. The new offer ensures that
the winner remains the winner regardless of any bid increment.

As mentioned earlier, limited by the MIP-based verification step, we are restricted
to tiny networks.
As described by the Lottery
Ticket Hypothesis~\cite{Frankle19:Lottery}, for tiny networks, having a ``lucky''
initialization is important. Motivated by this, the last monotonicity-seeking technique is simply {\em repeated trials}, which turns out to be effective for most distributions that we experimented on.
That is,
when we encounter either a poorly performing allocation or when monotonicity
verification fails, we simply train again from scratch with a fresh
initialization, until a near-optimal and verifiably monotone interpolation appears.
For many distributions, {\em repeated trials are all we need.}
On the other hand,
there are indeed distributions under which repeated trials do not work,
and we do need the full suite of techniques, including counterexample-guided training and the post-processing monotonicity fix,
to achieve near-optimal revenue.
In Appendix~\ref{subsec:worst10and100}, we describe $2$ adversarial distributions generated via a separate MIP-based heuristic.
In Appendix~\ref{sec:100trials}, we conducted $100$ training trials for each of these adversarial distributions, corresponding to {\em hundreds of hours} of computation.
While we do manage to get verifiably monotone allocations via (now a lot more) repeated trials, the achieved revenue is much worse
than the revenue obtained using the full suite of techniques, which is near-optimal.
That is, for the two adversarial distributions, even by trying for hundreds of hours, we still cannot reach a near-optimal auction.
This demonstrates the need for counterexample-guided training and the monotonicity fix.
Lastly, the monotonicity fix offers the peace of mind that regardless of the distribution, our approach always delivers strategy-proofness.

\vspace{.05in}
\noindent {\bf Contribution 3:
Extensive experiments demonstrating near-optimal revenue across all distributions tested, with substantial improvements over baselines.}

We performed an extensive suite of experiments on $65$ correlated distributions, ranging from synthetic (randomly generated and hand-crafted) and real-world data spanning consumer, financial, and industrial domains (Appendix~\ref{sec:realdata}), to adversarial distributions generated via two distinct methods to stress test our technique (Appendix~\ref{sec:adversarial}). Our performance is impressive across the board in terms of the revenue gap from the (unattainable) greedy upper bound. For instance, in Table~\ref{tab:n2-0}, the best baseline---the fully strategy-proof variant of RegretNet~\cite{Dutting19:Optimal}, which is also based on our techniques---has an average revenue gap of $2.7\%$, while our neural network interpolation method achieves an average revenue gap of only $0.26\%$, representing a tenfold improvement!

Optimal auction design under correlation is a foundational challenge in mechanism design. We provide a robust empirical resolution to this problem through extensive evaluation. By introducing the (unattainable) greedy revenue as a rigorous upper bound -- the first such benchmark for this setting -- we demonstrate that our method consistently achieves near-optimal revenue. This evaluation is supported by a full suite of ablation studies and exhaustive comparisons against all established manual and neural baselines. We provide the complete breadth of these results and technical specifications in the Appendix, alongside our implementation code in the supplementary materials, to ensure total transparency and reproducibility.

\vspace{.05in}
\noindent {\bf Contribution 4: Examples demonstrating some generalisation capabilities of our techniques.}

In Appendix~\ref{sec:multi}, we extend our techniques to {\em multi-unit auctions with unit demand}, showing similarly strong performance.
In Appendix~\ref{sec:rnet}, we show that it is convenient to integrate our techniques into RegretNet~\cite{Dutting19:Optimal} to design fully strategy-proof auctions,
while the original RegretNet is only approximately strategy-proof.
RegretNet involves both the allocation and the payment networks.
We require the allocation network be a tiny MLP with ReLU activation, so that we can apply
our MIP-based techniques, including verification and post-processing monotonicity fix, to ensure that the final allocation is monotone.
There is no restriction on the size or architecture of the payment network, as the payment network
only serves as a surrogate, which will be thrown away when training ends.
As long as the allocation is monotone, the ``correct'' payments can be reverse engineered
from the allocation following Myerson's characterization.

\section{Model Description}
\label{sec:model}

Following Myerson's setting, we aim to design revenue-maximizing single-item auctions that are {\em deterministic}, {\em strategy-proof} and
{\em ex post individually rational}. There are $n$ bidders. Bidder $i$'s valuation for the single item
is denoted as $b_i$. Without loss of generality, we assume $0\le b_i\le 1$. Since we focus on strategy-proof auctions,
we do not differentiate between reported bids and private valuations.
We use $\vec{b}=(b_1,b_2,\ldots,b_n)$ to denote the bid profile, which is drawn from a correlated distribution
with the joint probability density function $\phi(\vec{b})$. We assume $\phi$ is continuous, everywhere positive,
and bounded above by a constant.
Our goal is to design an auction that maximizes expected revenue under $\phi$.

Following \cite{Myerson81:Optimal,Papadimitriou11:On}, all deterministic, strategy-proof,
and ex post individually rational mechanisms for our setting can be interpreted as {\em price-oriented rationing-free}
mechanisms~\cite{Yokoo03:Characterization}. That is, every bidder faces a deterministic take-it-or-leave-it offer,
which potentially depends on the other bids.
The offers must be structured so that for each bid profile,
at most one bidder is willing to accept her offer.\footnote{Considering that our objective is revenue in expectation and we assume a continuous probability density function bounded above by a constant, we can ignore all tie-breaking issues.}

An allocation function $a$ maps each bid profile $\vec{b}$ to a binary vector
$(a_1,a_2,\ldots,a_{n+1})$, where $a_i$ being $1$ means bidder
$i$ wins. The last dimension $a_{n+1}$ represents an auxiliary bidder, whose
win results in the item not allocated. We require $\sum_i a_i=1$.  We use $a(\vec{b})_i$
to denote the $i$-th dimension of the allocation vector. Allocation
monotonicity requires that $\forall i,b_{-i},b_i,b_i'$ with $b_i<b_i'$,
$a(b_i,b_{-i})_i \le a(b_i', b_{-i})_i$.
Myerson's characterization says that the allocation rule uniquely determines the payment rule.
Using the take-it-or-leave-it offer interpretation, if bidder $i$ wins under
profile $(b_i,b_{-i})$, then her payment must be exactly $\inf\{b_i'|a(b_i',b_{-i})_i=1\}$.
Losing bidders pay nothing.

Figure~\ref{fig:mainfigure} illustrates an example monotone allocation for 2
bidders. Every bid profile $(b_1,b_2)=(x,y)$ corresponds to a point in the unit square.
Region A, i.e., the
blue points, is where bidder $1$ wins. Region B, i.e., the red points, is
where bidder $2$ wins. The white region is where the item is not allocated.
Allocation monotonicity can be interpreted as follows:
the blue region must be {\em rightward-closed}: for every blue
point, all points on its right must also be blue.  Similarly, the red region
must be {\em upward-closed}: for every red point, all points above
it must be red.  The region boundaries characterize the payments.
If $(x,y)$ is white, then the item is not allocated and no bidders pay.  If
$(x,y)$ is blue, then bidder $1$ wins and her payment is the
minimum $x'$ value so that $(x',y)$ is on the blue boundary.  If $(x,y)$ is red, then
bidder $2$ wins and his payment is the minimum $y'$ value so that
$(x,y')$ is on the red boundary.

\section{Virtual Valuation, Ironed Virtual Valuation, and Marginal Profit}
\label{sec:vv}

\cite{Myerson81:Optimal} introduced the {\em virtual valuation} and also a mathematical maneuver called {\em ironing}, leading
to the {\em ironed virtual valuation}.
\cite{Papadimitriou11:On} introduced another variant called {\em marginal profit}, which comes with a technical change that is helpful for our neural interpolation approach, which we will discuss later in this section.
While we can define the greedy allocation based on any of these three variants,
the marginal profit version consistently comes out on top in experiments, as evidenced
by Table~\ref{tab:n2-0}.
Due to space constraint, in this section, we only provide the bare minimum coverage of these concepts.
Details are deferred to Appendix~\ref{sec:appendixvv}.
Exposition of the virtual valuations, including alternative interpretations, can also be found in \cite{Bulow89:simple,Fu17:Notes,Hartline13:Mechanism}.

{\bf Virtual valuation:}
The virtual valuation is defined for independent bidders.
Bidder $i$'s virtual valuation only depends on her own distribution and her own bid.
We use $f_i$ and $F_i$ to denote the probability density function and the cumulative
distribution function of bidder $i$'s valuation.
We use $o$ to denote the
take-it-or-leave-it offer bidder $i$ faces,
with the understanding that $o$
depends on the other bids $b_{-i}$ (but this turns out to be irrelevant in the mathematical analysis).
Since it is a take-it-or-leave-it auction
from the perspective of bidder $i$, the expected revenue we can extract from $i$ is
$o(1-F_i(o))$.
The derivative of $o(1-F_i(o))$ with respect to $o$ equals $1-F_i(o)-of_i(o)=-(o-\frac{1-F_i(o)}{f_i(o)})f_i(o)$.
Since $F_i(1)=1$, we can show
\begin{equation}
    o(1-F_i(o))
=\int_{o}^1\left(x-\frac{1-F_i(x)}{f_i(x)}\right)f_i(x)dx
    \label{eq:vv}
\end{equation}
The left side of the formula is the expected revenue we can extract from $i$.
The right side can be interpreted as follows: Consider bidder $i$ who bids $x$. When $i$ wins, which is when
$x\ge o$, her contribution to the revenue is
$v_i(x)=x-\frac{1-F_i(x)}{f_i(x)}$, which is called $i$'s virtual
valuation.
When $i$ loses, which is when $x\le o$, her contribution to the
revenue is $0$ as $x$ is below the integration lower limit.  Bidder $i$'s expected
contribution in terms of virtual valuation equals the
expected revenue from $i$.  This interpretation leads to Myerson's greedy
allocation. That is, we simply convert the bidders' original bids to their
virtual valuations according to the above function $v_i$, and allocate the item to whoever has the highest virtual
valuation. If the highest virtual valuation is negative,
then we do not allocate.
Being greedy, this allocation goes for the maximum
revenue.  The caveat is that it may not be
monotone.  Myerson
calls a distribution {\em regular}
if under it, $v_i$ is nondecreasing.  If all bidders'
distributions are regular and independent, then it is easy to show that the greedy allocation
is monotone. If the winner increases her bid while the other bids stay the same, then due to distribution
regularity, the winner's virtual valuation never decreases.
The other bidders' virtual valuations
do not change due to the independence assumption, so the winner still wins when she raises her bid while the other bids stay the same.
The virtual valuation can be extended to correlated distributions by switching
to conditional probability functions.

{\bf Ironed virtual valuation:}
Above we illustrated that monotonicity holds assuming both distribution regularity (i.e., nondecreasing virtual valuations)
and independence.
Myerson proposed an ironing technique that
converts a not necessarily nondecreasing
virtual valuation function to a nondecreasing version.
Unfortunately, this ironing process is only mathematically sound with the independence assumption.
Considering this limitation,
we defer the details of the ironing process
to Appendix~\ref{sec:appendixvv}.

{\bf Marginal profit:}
The final variant of virtual valuation is marginal profit~\cite{Papadimitriou11:On}.
Our presentation of the marginal profit is slightly modified from the original, mainly for simplifying presentation (refer to Appendix~\ref{sec:appendixvv} for the original
definition).
We still use $f_i$ and $F_i$ to denote the probability density function
and the cumulative distribution function of bidder $i$.
The key difference between marginal profit and virtual valuation lies in
the following {\em revenue fix} process.

{\bf Revenue fix:}
Given a monotone allocation, for any bidder $i$ and any set of other bids $b_{-i}$,
$i$ faces a take-it-or-leave-it offer $o$, which depends on $b_{-i}$.
We can raise the value
of $o$ arbitrarily and this would never break monotonicity or cause over-allocation.
Therefore, suppose we have an allocation that offers $o$ to bidder $i$ when the other bids are $b_{-i}$, instead of offering $o$ directly and achieve a revenue of $o(1-F_i(o))$, we should {\em optimally} raise the offer to $\arg\max_{x\ge o}x(1-F_i(x))$,
leading to a revenue of $\max_{x\ge o}x(1-F_i(x))$.

The following equation defines the marginal profit, which is similar to Equation~\ref{eq:vv} for virtual valuation.
In Equation~\ref{eq:vv}, the left-hand side is the expected revenue extracted from bidder $i$.
In Equation~\ref{eq:mp} below, the left-hand side is the expected revenue extracted from bidder $i$, {\em after the revenue fix}.
\begin{equation}
    \label{eq:mp}
\max_{x\ge o}x(1-F_i(x)) =
    \int_o^1
\left(-\frac{\partial(\max_{x'\ge x}x'(1-F_i(x')))}{\partial x}/f_i(x)\right)f_i(x)dx
\end{equation}
We define $m_i(x)=-\frac{\partial(\max_{x'\ge x}x'(1-F_i(x')))}{\partial x}/f_i(x)$ to be the marginal profit
of bidder $i$ when her bid is $x$.
The message of Equation~\ref{eq:vv} is that the expected revenue is equal to the expected virtual valuation.
Equation~\ref{eq:mp} conveys a similar message, which is that the expected revenue {\em after the revenue
fix} is equal to the expected marginal profit.

{\bf Virtual valuation versus marginal profit:}
Although the two concepts are minor variations of each other, interpolating based on marginal profit offers critical advantages, as confirmed by experiments in Table~\ref{tab:n2-0}. A detailed technical comparison is provided in Appendix~\ref{sec:appendixvv}.  Here, we briefly discuss the high-level benefits of interpolating based on marginal profit. We recall that the greedy allocation based on the original virtual valuations works as follows: we allocate the item to the agent with the highest virtual valuation, {\em and if all agents' virtual valuations are negative, then we do not allocate}. On the other hand, marginal profits are always non-negative by construction. That is, the greedy allocation based on marginal profits simply allocates the item to the agent with the highest marginal profit, eliminating the need to explicitly consider the ``no allocation'' case. If all agents' marginal profits are zero, then {\em any} allocation is optimal. This simplification is possible because marginal profits are defined with the {\em revenue fix} step in mind. That is, the greedy allocation based on
marginal profits can defer the ``when not to allocate'' decisions to the revenue fix step.

The following example illustrates the revenue-fix process and shows how interpolating based on marginal profits effectively allows more ``slack'' for the neural interpolation.

{\bf Revenue fix example:} We refer to Figure~\ref{fig:greedy}, which shows an example greedy allocation that is not monotone.
Blue points are when bidder $1$'s marginal profits are strictly higher.
Red points are when bidder $2$'s marginal profits are strictly higher.
White points are when both bidders' marginal profits are zero.
We train a neural network to fit the greedy allocation, resulting in a monotone allocation depicted in Figure~\ref{fig:learnt}.
Since our network size is tiny, it glosses over the fine details, thereby avoiding the minor monotonicity violations
by the greedy allocation (i.e., the blue points enclosed by the red region and the red points enclosed by the blue region).
We note that when a point's marginal profit is $0$, it does not matter which bidder wins. In Figure~\ref{fig:learnt}, the network simply allocates the white points arbitrarily. That is, the focus of interpolation is on regions that actually matter (i.e.,
the limited expressive power of our tiny network focuses on creating a separation between A and B).
Figure~\ref{fig:revenuefix} shows the aftermath of the revenue
fix. The learned allocation allocates A+C to bidder $1$ and B+D to bidder $2$. After the fix, C and D become unallocated.
The final allocation is Figure~\ref{fig:mainfigure}.

\section{Details of the Proposed Approach}
\label{sec:technical}

We train a neural network to mimic the greedy allocation, which has three
versions, depending on whether we greedily allocate
according to the vanilla virtual valuation, the ironed virtual valuation, or the
marginal profit.
When there is no ambiguity, greedy allocation refers to the version based on marginal profit.
Even though we are maximizing revenue, we do not
need to reference any payment function in training.
Training is solely carried out on the allocation
function.

We represent the allocation rule as an MLP with ReLU activation. The inputs are the bids ($n$ dimensions).
The output dimension is $n+1$. If the $i$-th ($i=1,2,\ldots, n$) output coordinate is the highest, then bidder
$i$ wins. If the $(n+1)$-th coordinate is the highest, then the item is not allocated.

In Myerson's original approach, during the derivation of the optimal auction, randomization is allowed, which
simplifies the mathematical analysis.
Similarly, we allow randomization to facilitate the training.
We apply softmax to the network's outputs, ensuring that the coordinate with the highest value corresponds to the highest proportion of the item won in the context of randomized auctions. %
During training, we sample a batch of bid profiles. For each profile $\vec{b}$,
we calculate the marginal profits of each bidder. The marginal profit of the auxiliary bidder $(n+1)$ is always $0$.
We use $m(\vec{b})$ to represent the vector of marginal profits.
We use \textit{NN} to represent the network.
The training loss is simply the batched sum of $-m(\vec{b}) \cdot \text{softmax}(\mathit{NN}(\vec{b}))$ over all training
samples in the batch.
This mimics greedy allocation because in the case of perfect fit, if the marginal profit of bidder $i$ is the highest, then the $i$-th coordinate of $\text{softmax}(\mathit{NN}(\vec{b}))$ should be $1$ and the other coordinates should be $0$'s.
During evaluation, we revert back to the deterministic version. That is, we do not apply softmax and simply pick the highest output coordinate as the winner.

Our techniques also include the use of two mixed-integer programs. The first is for verifying
whether the trained allocation is monotone. For a trained network, its
weights and biases are viewed as constants. Since our network is
based on MLP+ReLU, for any node in the network, its value before activation can
be written as a linear expression involving the activated versions of the node values
from the previous layer. The ReLU activation step can be modelled using the
big-M trick with the introduction of an auxiliary binary variable.
When presenting the mixed-integer programs, we simply use $(a_1,a_2,\ldots,a_{n+1}) = \mathit{NN}(\vec{b})$ to
represent that the $a_i$'s are the outputs of the network when the input is $\vec{b}$, with the understanding that the $a_i$ can be written as linear expressions of $\vec{b}$ with
the help of auxiliary binary variables.
The MIP for monotonicity verification finds the largest gap between $b_i$
and $b_i'$ so that $i$ wins when bidding $b_i$, but after increasing
her bid to $b_i'$ while keeping the other bids fixed, a different bidder $j$ wins ($j$ can be $n+1$). The gap is the maximum
over $\forall b_{-i}$. In order to verify monotonicity, we need
to run $n^2$ MIPs (for every $i$ from $1$ to $n$ and for every $j\neq i$ from $1$ to $n+1$). All MIPs' objectives must be $0$'s (or infeasible) to conclude monotonicity.

\begin{minipage}{0.5\textwidth}
\begin{center}
{\bf MIP for monotonicity verification}
\begin{align*}
    \text{Constants} \quad & 1\le i\le n; j\neq i; 1\le j\le n+1\\
    \text{Variables} \quad & 0\le b_1,b_2,\ldots,b_n, b_i' \le 1;b_i \le b_i'\\
\text{Maximize} \quad & b_i'-b_i \\
    \text{subject to} \quad & (a_1,a_2,\ldots,a_{n+1}) = \mathit{NN}(b_i,b_{-i})\\
                & a_i = \max_t a_t \\
                & (a_1',a_2',\ldots,a_{n+1}') = \mathit{NN}(b_i',b_{-i})\\
                & a_j' = \max_t a_t' \\
\end{align*}
\end{center}
\end{minipage}
\hfill  %
\begin{minipage}{0.5\textwidth}
\begin{center}
{\bf MIP for monotonicity fix}
\begin{align*}
    \text{Constants} \quad & 1\le i\le n; j\neq i; 1\le j\le n+1\\
                           & 0\le b_1,\ldots,b_{i-1},b_{i+1},\ldots,b_n \le 1\\
    \text{Variables} \quad & 0\le b_i' \le 1\\
\text{Maximize} \quad & b_i' \\
    \text{subject to} \quad & (a_1,a_2,\ldots,a_{n+1}) = \mathit{NN}(b_i',b_{-i})\\
                & a_j = \max_t a_t \\
                &\\
                &\\
\end{align*}
\end{center}
\end{minipage}

The by-products of the MIP for monotonicity verification are the counterexamples.
Given a counterexample, we add a penalty term
$\lambda \cdot \text{ReLU}(\text{softmax}\left(\mathit{NN}(b_i,b_{-i}))_i-\text{softmax}(\mathit{NN}(b_i',b_{-i}))_i\right)$,
which aims to ensure that by increasing her bid, bidder $i$'s proportion of item won must never decrease.

With additional training on counterexamples, we sometimes, but not always, end up with verifiably monotone allocations.
When we do not manage to achieve verifiable monotonicity, we can always apply a post-processing
monotonicity fix using the second MIP. This fix is applied after collecting the bids. When
calculating $i$'s fix, the bids in $b_{-i}$ are treated as constants. For bidder $i$, the second MIP finds
the maximum bid for $i$ where $i$ loses (according to the trained network) to a different bidder $j$ ($j$ could be $n+1$).
We need to solve $n$ MIPs only for the (tentatively, i.e., according to the trained network) winning bidder, and the maximum over the $n$ objective values (ignoring infeasible MIPs) is
the monotonicity cutoff price for $i$. As $i$'s bid goes from this cutoff price to $1$, $i$ always wins, which
implies monotonicity.  But if $i$'s bid was below this, $i$ does not win after all.
The monotonicity fix is applied together with the revenue fix from Section~\ref{sec:vv}.
For example, let the original offer produced by the network be $o$. The revenue fix instructs to
raise the offer to $o_r$ and the monotonicity fix instructs to raise the offer to $o_m$.
The final offer is then $\max\{o,o_r,o_m\}$.

\section{Main Experiments}
\label{sec:exp}

\begin{wrapfigure}{r}{0.25\textwidth}  %
    \begin{minipage}{0.25\textwidth}
\vspace{-.7in}
    \begin{footnotesize}
\begin{equation*}
\begin{bmatrix}
11 & 24 & 7 & 20 & 3 \\ 4 & 12 & 25 & 8 & 16 \\ 17 & 5 & 13 & 21 & 9 \\ 10 & 18 & 1 & 14 & 22 \\ 23 & 6 & 19 & 2 & 15
\end{bmatrix}\\
\end{equation*}
    \end{footnotesize}

    \end{minipage}
\end{wrapfigure}

To facilitate experiments, we introduce the {\em grid distributions}, where we
use $n$-dimensional matrices to represent correlated distributions for $n$
bidders. The example 2D matrix below represents a correlated
distribution for $2$ bidders.  Here, the matrix size is $5\times 5$.  We divide
the unit square $[0,1]\times [0,1]$ into $25$ sub-squares of side length
$\frac{1}{5}$.  The bottom left element is $23$, which means that when the bid
profile $(x,y)$ falls into sub-square $[0,\frac{1}{5}]\times
[0,\frac{1}{5}]$, the joint probability density is {\em uniformly} $23$. The
element on the right of $23$ is $6$, which means that for the sub-square
$[\frac{1}{5},\frac{2}{5}]\times [0,\frac{1}{5}]$, the joint probability
density is uniformly $6$.\footnote{There is an additional normalization ratio, which will be multiplied with every density value to ensure that
the total probability equals exactly $1$. It is omitted to make the presentation cleaner.}

There are multiple reasons for introducing the grid distributions. 1) They allow the {\em systematic generation} of correlated
distributions. 2) All relevant numerical operations can be implemented {\em analytically}, including derivative, maximum, and integration. This is important for producing {\em numerically stable} experimental results as the marginal profit is the
{\em partial derivative} of the {\em maximum} of a term involving the cumulative distribution function, which calls for an {\em integration}. 3) Theoretically speaking, the grid distributions can
approximate {\em any} correlated distribution if we allow arbitrarily fine grids. 4) The grid distribution family
makes it easy to generate {\em adversarial distributions} via tuning the matrix.
For example, in Appendix~\ref{sec:adversarial}, we report on experiments using evolutionary computation to evolve an adversarial matrix, as well as
generating adversarial matrices using mixed-integer-programming. 5)
We can easily construct hand-crafted grid distributions.
For example, we use {\em magic squares} -- 2D matrices in which every row and every column sums to the same value -- to create
correlated distributions where each bidder's marginal distribution is uniform.
The matrix shown above is a magic square.
We use magic-square-based distributions to evaluate the effect of correlation on revenue.
For such distributions, if we ignore correlation, then the distribution becomes i.i.d. uniform, where the second
price auction with reserve is optimal.
As a result, on correlated distributions, the gap between our revenue and the revenue from a second-price auction with a reserve price based on an i.i.d. distribution represents the additional revenue we can gain by considering correlation. %

\begin{table*}[h!]
\centering
    \caption{$2$ bidders: \begin{footnotesize}$\textsc{Gap}(\textsc{MP})=0.26\%$,$\textsc{Gap}(\textsc{RNet})=2.7\%$,$\textsc{Gap}(\textsc{2nd})=4.2\%$,$\textsc{Gap}(\textsc{Myer})=4.3\%$\end{footnotesize}
    }
\label{tab:n2-0}
\begin{footnotesize}
\begin{tabular}{lll|llllll}
\toprule
{\sc Distrib.} &  {\sc Greedy} & {\sc AMD} & {\sc Myer.} & {\sc 2nd} & {\sc RNet} & {\sc MP} & {\sc VV} & {\sc IVV} \\
\midrule

$G5_0$ & $0.4353$ & $0.4368$ & $0.4191$ & $0.4197$ & $0.4307$ & $\textbf{0.4368}$ & $0.4355$ & $0.4329$\\
$G5_1$ & $0.4047$ & $0.4041$ & $0.3939$ & $0.3929$ & $0.4021$ & $\textbf{0.4037}$ & $0.4032$ & $0.4032$\\
$G5_2$ & $0.3979$ & $0.3967$ & $0.3771$ & $0.3823$ & $0.3869$ & $\textbf{0.3959}$ & $0.3946$ & $0.3955$\\
$G5_3$ & $0.4629$ & $0.4625$ & $0.4533$ & $0.4546$ & $0.4522$ & $\textbf{0.4628}^*$ & $0.4612$ & $0.4616$\\
$G5_4$ & $0.4681$ & $0.4665$ & $0.4476$ & $0.4472$ & $0.4639$ & $\textbf{0.4674}^*$ & $0.4673$ & $0.4671$\\
$G5_5$ & $0.4204$ & $0.4187$ & $0.4075$ & $0.4057$ & $0.4062$ & $\textbf{0.4151}$ & $0.4148$ & $0.4150$\\
$G5_6$ & $0.3905$ & $0.3912$ & $0.3761$ & $0.3741$ & $0.3798$ & $\textbf{0.3908}$ & $0.3893$ & $0.3886$\\
$G5_7$ & $0.4865$ & $0.4854$ & $0.4494$ & $0.4544$ & $0.4771$ & $\textbf{0.4850}$ & $0.4828$ & $0.4840$\\
$G5_8$ & $0.4298$ & $0.4273$ & $0.3988$ & $0.3981$ & $0.4077$ & $\textbf{0.4291}^*$ & $0.4281$ & $0.4279$\\
$G5_9$ & $0.4531$ & $0.4512$ & $0.4306$ & $0.4302$ & $0.4379$ & $\textbf{0.4498}$ & $0.4482$ & $0.4497$\\
{\sc saturn} & $0.4533$ & $0.4545$ & $0.4210$ & $0.4226$ & $0.4372$ & $0.4510$ & $0.4512$ & $\textbf{0.4515}$\\
{\sc jupiter} & $0.4427$ & $0.4404$ & $0.4238$ & $0.4219$ & $0.4316$ & $\textbf{0.4418}^*$ & $0.4409$ & $0.4410$\\
{\sc mars} & $0.4375$ & $0.4377$ & $0.4177$ & $0.4167$ & $0.4194$ & $\textbf{0.4360}$ & $0.4333$ & $0.4359$\\
{\sc sol} & $0.4429$ & $0.4414$ & $0.4376$ & $0.4364$ & $0.4329$ & $\textbf{0.4427}^*$ & $0.4404$ & $0.4419$\\
{\sc venus} & $0.4381$ & $0.4381$ & $0.4175$ & $0.4171$ & $0.4233$ & $\textbf{0.4365}$ & $0.4319$ & $0.4358$\\
{\sc mercury} & $0.4273$ & $0.4274$ & $0.4179$ & $0.4176$ & $0.4150$ & $\textbf{0.4281}^*$ & $0.4271$ & $0.4279$\\
{\sc luna} & $0.4331$ & $0.4335$ & $0.4173$ & $0.4172$ & $0.4212$ & $\textbf{0.4322}$ & $0.4315$ & $0.4318$\\

\bottomrule
\end{tabular}
\end{footnotesize}
\end{table*}

\begin{table*}[h]
    \centering
    \caption{$3$ and $5$ bidders:
    \begin{footnotesize}
    {\sc M/2} represents the
    better between {\sc Myerson} and {\sc 2nd}\\
    $n=3$: $\textsc{Gap}(\textsc{MP})=0.36\%$,
    $\textsc{Gap}(\textsc{RNet})=3.6\%$,
    $\textsc{Gap}(\textsc{2nd})=3.9\%$,
    $\textsc{Gap}(\textsc{Myer})=3.8\%$\\
    $n=5$: $\textsc{Gap}(\textsc{MP})=0.87\%$,
            $\textsc{Gap}(\textsc{RNet})=2.8\%$,
    $\textsc{Gap}(\textsc{2nd})=2.4\%$,
    $\textsc{Gap}(\textsc{Myer})=2.4\%$
    \end{footnotesize}
    }
  \label{tab:n35}
  \centering
    \begin{footnotesize}
  \begin{tabular}{llllllllll}
    \toprule
      & \multicolumn{4}{c}{$n=3$} && \multicolumn{4}{c}{$n=5$} \\
     \cmidrule(r){2-5}
     \cmidrule(r){7-10}
      {\sc Distrib.} &  {\sc Greedy} & {\sc M/2} & {\sc RNet} & {\sc MP} & & {\sc Greedy} & {\sc M/2} & {\sc RNet} & {\sc MP} \\
    \midrule

$G5_0$ & $0.5512$ & $0.5346$ & $0.5346$ & $\textbf{0.5517}$
      &&
$0.6862$ & $0.6700$ & $0.6670$ & $\textbf{0.6799}$\\

$G5_1$ & $0.5647$ & $0.5475$ & $0.5509$ & $\textbf{0.5651}$
      &&
$0.6901$ & $0.6738$ & $0.6713$ & $\textbf{0.6846}$\\

$G5_2$ & $0.5316$ & $0.5137$ & $0.5153$ & $\textbf{0.5302}$
      &&
$0.6838$ & $0.6709$ & $0.6681$ & $\textbf{0.6783}$\\

$G5_3$ & $0.5476$ & $0.5283$ & $0.5290$ & $\textbf{0.5450}$
      &&
$0.6894$ & $0.6724$ & $0.6684$ & $\textbf{0.6838}$\\

$G5_4$ & $0.5485$ & $0.5306$ & $0.5310$ & $\textbf{0.5471}$
      &&
$0.6853$ & $0.6692$ & $0.6664$ & $\textbf{0.6787}$\\

$G5_5$ & $0.5477$ & $0.5237$ & $0.5255$ & $\textbf{0.5443}$
      &&
$0.6850$ & $0.6720$ & $0.6685$ & $\textbf{0.6809}$\\

$G5_6$ & $0.5371$ & $0.5155$ & $0.5158$ & $\textbf{0.5332}$
      &&
$0.6902$ & $0.6725$ & $0.6686$ & $\textbf{0.6839}$\\

$G5_7$ & $0.5723$ & $0.5528$ & $0.5555$ & $\textbf{0.5710}$
      &&
$0.6900$ & $0.6719$ & $0.6682$ & $\textbf{0.6824}$\\

$G5_8$ & $0.5473$ & $0.5193$ & $0.5185$ & $\textbf{0.5429}$
      &&
$0.6868$ & $0.6715$ & $0.6659$ & $\textbf{0.6807}$\\

$G5_9$ & $0.5663$ & $0.5410$ & $0.5396$ & $\textbf{0.5637}$
      &&
$0.6927$ & $0.6768$ & $0.6711$ & $\textbf{0.6866}$\\

    \bottomrule
  \end{tabular}
    \end{footnotesize}
\end{table*}

We altogether experimented with $62$ different grid distributions as follows.
{\em In Appendix~\ref{sec:continuous}, we provide additional experiments on $3$ continuous distributions.}

\vspace{.05in}
\textbullet\,
{\bf $\textbf{40}$ randomly generated grid distributions:} For $n$ from $2$ to $5$, we generate $10$ random grid distributions of size
       $\underbrace{5\times \ldots\times 5}_n$,
        where every matrix element is drawn from uniform $0$ to $1$.
We use $G5_s$ to denote the distribution generated with seed $s$.

\vspace{.05in}
\textbullet\,
{\bf $\textbf{7}$ hand-crafted grid distributions:} These are based on magic squares described in Appendix~\ref{sec:magic}.

\vspace{.05in}
\textbullet\,
{\bf $\textbf{12}$ adversarially generated distributions:} $10$ were generated using evolutionary computation and $2$ were
        generated using mixed-integer-programming. We defer all discussion on adversarial distributions to Appendix~\ref{sec:adversarial}.
        {\em We use them to test the limit of our proposed approach.}

\vspace{.05in}
\textbullet\,
{\bf $\textbf{3}$ empirical distributions derived from real-world datasets:}
These datasets come from diverse real-world market domains: (i) Consumer (eBay),
involving layperson bidding for electronics; (ii) Financial (Ethereum), involving algorithmic bidding for block building rights; and (iii) Electricity Market (PJM), involving professional strategic bids from energy producers.
Discussions on these distributions are deferred to Appendix~\ref{sec:realdata}.
\vspace{.05in}

As mentioned in Section~\ref{sec:vv}, there are three variants of the virtual
valuation, leading to three greedy allocations that can be used as interpolation targets.  For $2$ bidders, we compare
the revenue achieved using our approach based on {\sc MP} (marginal profit), {\sc VV}
(vanilla virtual valuation) and {\sc IVV} (ironed virtual
valuation). {\sc MP} consistently comes out on top. For $3$ to $5$ bidders, we only focus on {\sc MP}.
Our experiments also involve the following baseline auctions:

\vspace{.05in}
\textbullet\,
        {\bf {\sc {\bf Greedy}} (based on marginal profit):}
        For distributions we experimented on, the greedy allocation
        is rarely monotone.
        {\sc Greedy} violates the key constraint and only serves as the revenue {\em upper bound}.

\vspace{.05in}
\textbullet\,
        {\bf {\sc {\bf AMD}}:}
We round the bids down to the nearest multiple of $0.01$ and apply the classical approach to automated mechanism design~\cite{Conitzer02:Complexity,conitzer2004self}, which is a mixed-integer-programming-based approach producing the optimal auction when we restrict to discrete bids. We use AMD as an alternative implementation to the FPTAS for 2 correlated bidders from \cite{Papadimitriou11:On}, which
        is also based on discrete bids.
        Both AMD and FPTAS are for $2$ bidders only: AMD does not scale beyond $2$ bidders for our setting and the FPTAS only applies to $2$ bidders (optimal correlated auction design becomes NP-hard to approximate for $3$ bidders).

\vspace{.05in}
\textbullet\,
        {\bf {\sc {\bf Myerson}}:} Myerson's optimal auction ignoring correlation.
        The bidders' ironed virtual valuations are based on {\em marginal} distributions (instead of {\em conditional} distributions).

\vspace{.05in}
\textbullet\,
        {\bf {\sc {\bf 2nd}}:} Second price auction with the optimal reserve.

\vspace{.05in}
\textbullet\,
        {\bf {\sc {\bf RNet}}:} Integration of verification to RegretNet~\cite{Dutting19:Optimal}, details in Appendix~\ref{sec:rnet}.
\vspace{.05in}

In Tables~\ref{tab:n2-0} and \ref{tab:n35}, we present the results for $n\in \{2,3,5\}$.
Full tables including standard errors, results for $n=4$, and results for adversarial distributions are deferred to Appendix~\ref{sec:fulltables}.
A value is bold if it is the best among the scalable approaches ({\sc AMD} is not in the comparison).
{\sc MP} consistently achieves the best revenue among the scalable approaches.
When {\sc MP} also outperforms {\sc AMD}, we mark the result with a `*'.
Sometimes {\sc MP}'s value appears to be higher than the upper bound {\sc Greedy}. This
is caused by Monte Carlo simulation
(see evaluation details in Appendix~\ref{sec:parameters}).

{\sc Greedy}'s revenue is an unattainable upper bound. We use $\textsc{Gap}(\textsc{Method})$ to denote the average revenue gap between {\sc Method} and {\sc Greedy}, formally defined as

\[\frac{\text{Average revenue of {\sc Greedy}}-\text{Average revenue of {\sc Method}}}{\text{Average revenue of {\sc Greedy}}}\]

In Table~\ref{tab:n2-0}, the best baseline {\sc RNet} -- the fully strategy-proof variant of RegretNet~\cite{Dutting19:Optimal}, which is also based on our techniques -- has an average revenue gap of $2.7\%$, while our method {\sc MP} achieves an average revenue gap of only $0.26\%$, representing a tenfold improvement!

In Appendix~\ref{sec:multi}, we extend to multi-unit auctions with unit demand, where
we achieved similar strong performance. In Table~\ref{tab:multi},
the baseline $(m+1)$-th price auction with the optimal reserve has an
average revenue gap of $5.1\%$, while our method {\sc MP} has an average
revenue gap of $0.67\%$.

\section{Related Research}
\label{sec:related}
\cite{Papadimitriou11:On} proposed the {\em marginal profit} (compared with virtual valuation in Appendix~\ref{sec:appendixvv}), gave an FPTAS for two correlated bidders, and showed that the problem is NP-hard to approximate for three or more.
\cite{Dobzinski11:Optimal} showed that under truthful-in-expectation, the optimal auction for correlated bidders can be found in time polynomial in the support size.
Randomization matters here: determinism is without loss for independent bidders~\cite{Myerson81:Optimal} but costs revenue under correlation~\cite{Caragiannis16:Limitations}.
\cite{Roughgarden13:Optimal} gave a sufficient condition for Myerson's greedy allocation to remain monotone under correlation, but it is unclear how to check the condition computationally for a given distribution, and for most distributions in this paper the greedy allocation is in fact not monotone.
\cite{Ronen01:On} proposed
an approximate auction that guarantees half of the revenue for correlated bidders.
\cite{Cremer88:Full} showed that it is possible to extract the full social surplus
for correlated bidders, but the proposed auction is only individually rational in expectation.
\cite{Bei19:Correlation-robust} studied single-item auctions for correlated bidders
from the lens of worst-case analysis. The authors evaluate an auction based on its revenue under the worst-case correlation.

This paper follows the line of research on neural network mechanism design
initiated by \cite{Dutting19:Optimal}. \cite{Curry20:Certifying} first
applied mixed-integer-programming to evaluate strategy-proofness of neural-network-based mechanisms.
\cite{Guo24:Worst-case} proposed a suite of
techniques on worst-case mechanism design via neural networks, including
worst-case counterexample-guided training where the worst-case profiles are obtained via mixed-integer-programming and using the Lottery Ticket Hypothesis~\cite{Frankle19:Lottery} to
guide the search for tiny networks that are trainable for worst-case training objectives.
\cite{Duan24:Scalable} studied {\em virtual valuation combinatorial auctions}~\cite{Sandholm15:Automated} via neural networks, where ``virtual valuation'' is a different concept from Myerson's.
The monotonicity fix in our paper post-processes the allocation network to ensure strategy-proofness. GemNet~\cite{Wang24:GemNet}
proposed a similar approach that post-processes the payment network to ensure menu compatibility ({\em i.e.}, preventing two bidders from
winning the same item). Their method discretizes the bids
and then extend to general bids via Lipschitz smoothness. In contrast, our technique operates directly on continuous
values.

Monotonic network structures such as min-max networks~\cite{Sill97:Monotonic,Milani-Fard16:Fast} enforce monotonicity of the network outputs, which differs from allocation monotonicity despite the shared name.
The work closest to ours is \cite{Huo25:Learning}, which extends MyersonNet~\cite{Dutting19:Optimal} to correlated bidders.
MyersonNet assigns every bidder an allocation priority value computed by a min-max network from her own bid only, so it is monotone by construction but cannot fully exploit bid correlation.
\cite{Huo25:Learning} instead make each priority value a multi-input min-max network that is nondecreasing in the bidder's own bid and nonincreasing in the others' (the {\em Lopomo assumptions}~\cite{Roughgarden13:Optimal}), so allocation monotonicity still holds by construction, and payments are derived analytically from the network structure.
The price of guaranteeing strategy-proofness by construction is expressiveness.
In Appendix~\ref{sec:altnet}, we give an example where no allocation satisfying the Lopomo assumptions can express the optimal auction, and show experimentally that min-max variants, even when combined with our training techniques, fall below the second-price auction with an optimal reserve price, on the distributions that we experimented on.
We instead use unconstrained MLPs and establish strategy-proofness by formal verification, and we map marginal profits directly to payments, keeping a direct link to Myerson's theory without additional architecture for computing payments.
A second difference is in evaluation.
\cite{Huo25:Learning} treat RegretNet's revenue as the gold standard to approach, and use the second-price auction without reserve as baseline.
Neither anchors the results to optimality: RegretNet is only approximately strategy-proof and comes with no optimality guarantee, and the second-price auction without reserve is far from optimal.
We instead measure against the greedy revenue, a rigorous upper bound on the optimal revenue, and against stronger baselines: the second-price auction with an optimal reserve, Myerson's auction based on marginal distributions (Appendix~\ref{sec:baseline}), and a fully strategy-proof RegretNet variant built with our verification technique.
In Table~\ref{tab:n2-0}, the strategy-proof RegretNet variant has an average revenue gap of $2.7\%$ from the upper bound, while our method achieves $0.26\%$. The original RegretNet's revenue is inflated by its strategy-proofness violations and is therefore not comparable to that of strategy-proof auctions; once strategy-proofness is enforced, even the hand-crafted baselines sometimes outperform it (Tables~\ref{tab:n4-1} and \ref{tab:n5-1}).

\begin{ack}
This work was supported with supercomputing resources provided by
the Phoenix HPC service at Adelaide University.
Jiayuan Liu and Vincent Conitzer thank the Cooperative AI Foundation, Macroscopic
Ventures (formerly Polaris Ventures / the Center for Emerging Risk Research) and Jaan
Tallinn's donor-advised fund at Founders Pledge for financial support.
\end{ack}

\bibliographystyle{plainnat}
\bibliography{extra}

\appendix
\setcounter{section}{0}
\setcounter{secnumdepth}{2}
\renewcommand{\thesection}{\Alph{section}}
\renewcommand\thesubsection{\thesection.\arabic{subsection}}

\section{More on Virtual Valuation, Ironed Virtual Valuation, and Marginal Profit}
\label{sec:appendixvv}

\begin{figure*}[h]
    \centering
    \begin{subfigure}[b]{0.25\textwidth}
        \centering
        \includegraphics[width=\textwidth]{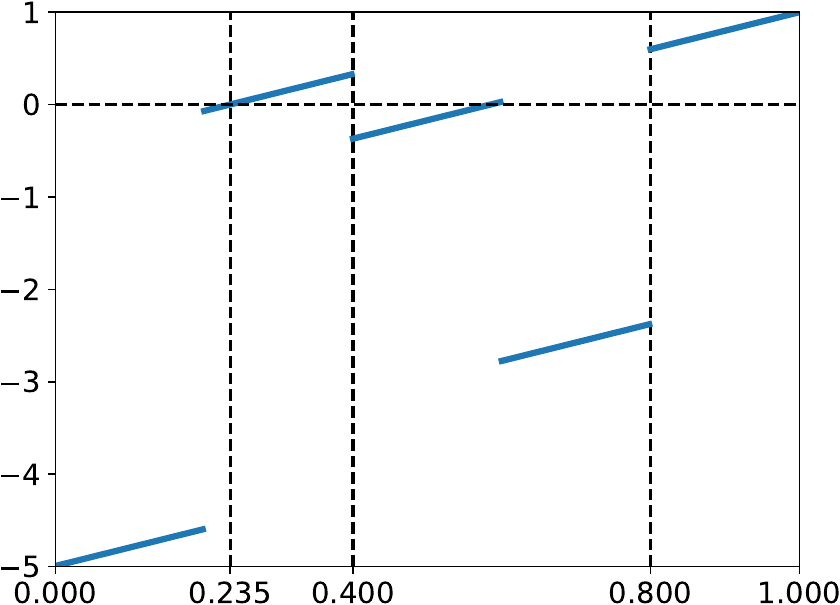}
        \caption{Virtual valuation}
        \label{fig:vv}
    \end{subfigure}
    \begin{subfigure}[b]{0.25\textwidth}
        \centering
        \includegraphics[width=\textwidth]{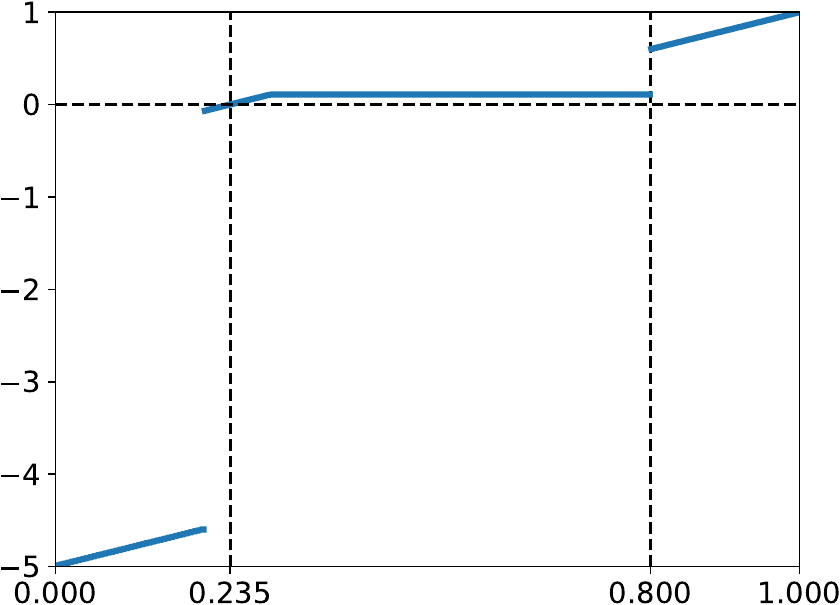}
        \caption{Ironed virtual valuation}
        \label{fig:ivv}
    \end{subfigure}
    \begin{subfigure}[b]{0.25\textwidth}
        \centering
        \includegraphics[width=\textwidth]{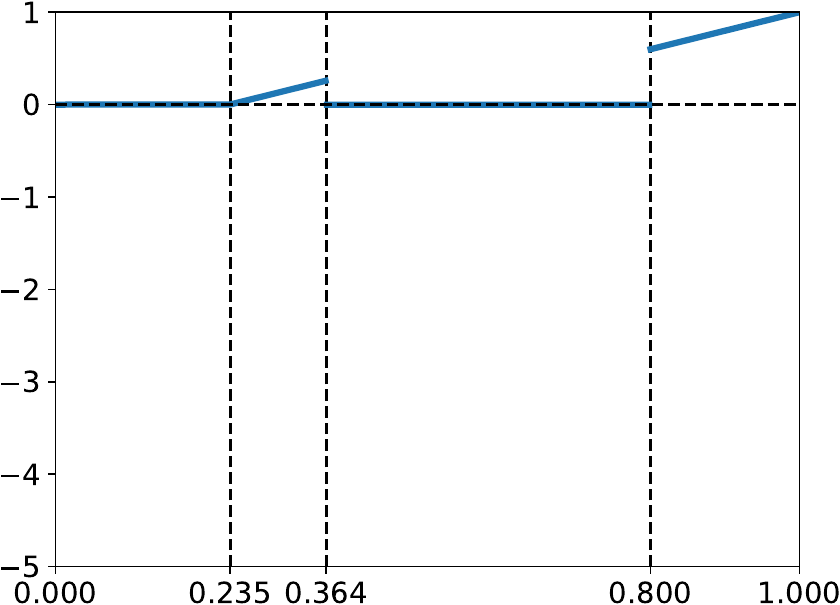}
        \caption{Marginal profit}
        \label{fig:mp}
    \end{subfigure}
    \caption{Plots for grid distribution $[0.2, 3.58, 0.32, 0.05, 0.85]$.
The grid distribution family is formally defined in Section~\ref{sec:exp}.
    The example grid distribution here describes a random variable with range $[0,1]$, which is divided into $5$ equally-sized sub-intervals. In $[0,\frac{1}{5}]$, the probability density value is uniformly $0.2$.
    In $[\frac{1}{5},\frac{2}{5}]$, the probability density value is uniformly $3.58$.}
    \label{fig:vvivvmp}
\end{figure*}

In this section, we present the omitted details from Section~\ref{sec:vv}.
We start with Figure~\ref{fig:vvivvmp}, which depicts the three variants of virtual
valuations for the example grid distribution $[0.2, 3.58, 0.32, 0.05, 0.85]$.

{\bf Ironed virtual valuation:}
The greedy allocation based on the virtual valuation guarantees monotonicity
when the distribution is both regular and independent.
Actually, the regularity assumption can be relaxed as long as independence holds.
Myerson proposed an ironing technique that
converts a not necessarily nondecreasing
virtual valuation function (i.e., Figure~\ref{fig:vv}) to a nondecreasing version (i.e., Figure~\ref{fig:ivv}).
The effect of ironing is that the decreasing regions (and some of their nearby regions)
become ``flat'', and within a flat region, the original virtual valuation is
replaced by the average (weighted according to the probability density).
Unfortunately, this averaging process is only mathematically sound with the independence assumption.
While we can still iron in the presence of correlation, and base the greedy allocation
on the ironed virtual valuations, the resulting auctions
are suboptimal, as confirmed by experiments in Table~\ref{tab:n2-0}.

Considering that ironing does not align well with the correlated setting, we only
very briefly describe its high-level idea and also
why it causes ``information loss'' in training.
We still use $f_i$
and $F_i$ to denote bidder $i$'s probability density function and cumulative distribution function.
Myerson's ironing approach requires the introduction of new notations.
We use $q=1-F_i(o)$ to denote bidder $i$'s {\em demand} when faced with offer $o$.  We use $R_i(q)$ to denote the revenue extracted from bidder $i$ when her demand is $q$.
That is, $R_i(q)=qo=qF_i^{-1}(1-q)$. The derivative of this revenue with respect to the demand is then $\frac{dR_i(q)}{dq}=
F_i^{-1}(1-q)-\frac{q}{f_i(F_i^{-1}(1-q))}$.
If we rewrite this derivative in terms of $o$, noting that $o=F_i^{-1}(1-q)$, then
the derivative becomes $o-\frac{1-F_i(o)}{f_i(o)}$, which is exactly $v_i(o)$, i.e., the virtual valuation at value $o$.
If the virtual valuation is nondecreasing in $o$, then it is nonincreasing in $q$, as higher offers correspond to lower demands.
This implies that
for regular distributions, $R_i(q)$ is {\em concave} (i.e., its derivative does not increase in $q$).
For non-regular distributions, $R_i(q)$ is not concave, which means that there exist two demand values $q_1$ and $q_2$,
so that $\frac{R_i(q_1)+R_i(q_2)}{2}\ge R_i(\frac{q_1+q_2}{2})$.
During Myerson's derivation of the optimal auction, randomized auctions are allowed. It is
just that the final optimal auction is proven to be deterministic. When randomized auctions are allowed,
the offer leading to demand $\frac{q_1+q_2}{2}$ is never a good idea, because more revenue can be achieved
by replacing this offer by a half/half mixture of the offers leading to demand $q_1$ and $q_2$.
In summary, Myerson showed that we can always resort to randomization to achieve a concave revenue function $\overline{R_i}(q)$
on the basis of the original $R_i(q)$, i.e., by pushing up $R_i(\frac{q_1+q_2}{2})$ to $\frac{R_i(q_1)+R_i(q_2)}{2}$ for
the non-concave regions.
In terms of implementation, one option is the Graham scan algorithm, a computational geometry algorithm for generating the convex hull given a set of points.
Recall that the derivative of the original revenue function $R_i(q)$ is the original virtual valuation.
Myerson uses the derivative of $\overline{R_i}(q)$ to serve as the ironed virtual valuation, which is guaranteed to be nondecreasing (in the bid) due to the concavity of $\overline{R_i}(q)$.
Note that $R_i$ and $\overline{R_i}$ would differ in one or several regions. For example, let $[q_1,q_2]$ be
one region where $R_i$ and $\overline{R_i}$ differ. The curve of $\overline{R_i}$ on $[q_1,q_2]$ is exactly
the straight line connecting $(q_1,R_i(q_1))$ and $(q_2,R_i(q_2))$, which means that the derivative of $\overline{R_i}$,
i.e., the ironed virtual valuation, must be a constant in $[q_1,q_2]$. This is how the flat region in Figure~\ref{fig:ivv}
is calculated.

Next, we discuss why the above ironing process is no longer valid for correlated
distributions and why it causes ``information loss'' in training.
In Figure~\ref{fig:ivv}, there is a long flat region $[0.289,0.8]$.
The virtual valuation in $[0.289,0.8]$ is replaced by the average $0.109$.
Suppose that the curve shown in Figure~\ref{fig:ivv} is for bidder $i$. For independent
bidders, when we apply the greedy allocation, if the other bidders' ironed
virtual valuations do not exceed $0.109$, then bidder $i$ would win if her bid
is at least $0.289$, which makes it correct to use the average to replace the
actual virtual valuations, because bidder $i$ ``wins'' for the whole interval
$[0.289,0.8]$. If the maximum of the other bidders' ironed virtual valuations exceeds $0.109$,
then bidder $i$ does not win with any bid below $0.8$, so the
virtual valuation changes in $[0.289,0.8]$ (due the averaging process) are irrelevant.
When we extend to correlated distributions (by defining ironed virtual valuations in terms of conditional distributions), the above analysis no longer holds. When bidder $i$ changes her bid within $[0.289,0.8]$, the other bidders' ironed virtual valuations may change due to correlation. For example, it is possible that when bidder
$i$ bids below $0.4$, there is another bidder whose virtual valuation is higher than $0.109$, and when bidder $i$ bids above
$0.4$, all other bidders' virtual valuations are below $0.109$. In this situation, within the flat region, bidder $i$ would only win in the region
$[0.4,0.8]$.
The ironed virtual valuation suggests that getting allocated in
$[0.4,0.8]$ is beneficial to the revenue, as this flat region is positive in Figure~\ref{fig:ivv}.
But according to the actual virtual valuation in Figure~\ref{fig:vv}, getting allocated in $[0.4,0.8]$ actually hurts the revenue.

{\bf Our presentation of marginal profit vs the original:}
Our presentation of marginal profit is slightly modified
from the original definition from \cite{Papadimitriou11:On}.
We use $m_i(b_i|b_{-i})$ to denote the marginal profit of bidder $i$ when her bid is $b_i$ and the other
bids are $b_{-i}$. Recall that $\phi$ is the joint probability density function.
We use $\varphi(b_{-i})$ to denote the joint probability density function of the bids except for $i$'s bid.
The original definition in \cite{Papadimitriou11:On} is
\[m_i(b_i|b_{-i})=-\frac{\partial\left(\max_{x'\ge b_i}x'\int_{x'}^1\phi(t,b_{-i})dt\right)}{\partial b_i}\]
\[=-\phi(\vec{b})\frac{\partial\left(\max_{x'\ge b_i}x'(1-F_i(x'|b_{-i}))\right)}{\partial b_i}\Big/f_i(b_i|b_{-i})\]

Our presentation (after switching to conditional distributions) is essentially

\[m_i(b_i|b_{-i})=-\frac{\partial\left(\max_{x'\ge b_i}x'(1-F_i(x'|b_{-i}))\right)}{\partial b_i}\Big/f_i(b_i|b_{-i}).\]

The only difference is the multiplier $\phi(\vec{b})$.
Since what we truly care about is the relative order of the marginal profits, i.e., which
bidder has the highest marginal profit, the two presentations are equivalent for this purpose.
Our presentation enabled the direct comparison between the marginal profit and the (ironed) virtual valuation in Figure~\ref{fig:vvivvmp}.

{\bf Detailed comparison between virtual valuation versus marginal profit:}
If the distribution is regular, then the derivative of $x(1-F_i(x))$ with respect to $x$, i.e., the original virtual valuation, is nondecreasing,
which means that there is a cutoff value $x^0$, so that $x(1-F_i(x))$ is nonincreasing when $x\le x^0$ and it is
nondecreasing when $x\ge x^0$. This implies $m_i(x)=0$ if $x\le x^0$ and $m_i(x)=v_i(x)$ if $x\ge x^0$.
That is, for regular distributions, $m_i(x)=\max\{v_i(x),0\}$. The only difference between the virtual valuation function $v_i$ and the marginal profit function $m_i$ is the additional ``max''.
For general distributions, the difference is more notable, but it remains based on the same underlying principle.

Although the two concepts are minor variations of each other, interpolating based on marginal profit offers critical advantages, as confirmed by experiments in Table~\ref{tab:n2-0}. Below, we provide an example to illustrate the benefits of marginal profit over virtual valuation.
We take the uniform distribution over $[0,1]$ as an example, which is regular. Suppose we only have one bidder. The optimal offer should be $0.5$, which can be interpreted as an optimal reserve price. Myerson's virtual valuation function is $v_i(x)=x-\frac{1-F_i(x)}{f_i(x)}=2x-1$.
The marginal profit function is $m_i(x)=\max\{2x-1,0\}$.
According to the virtual valuation, when the bid $x$ is below $0.5$, the virtual valuation is negative, which means
we should allocate the item only if the bid is at least $0.5$, therefore explaining the optimal reserve at $0.5$. On the other hand, the marginal profit is $0$
when the bid $x$ is below $0.5$. Therefore, when the bid is lower than $0.5$, allocating is neither helpful nor detrimental by this measure.
When the bid is higher than $0.5$, the marginal profit is positive, so allocating is helpful.
So according to marginal profit, every offer in $[0,0.5]$ is optimal, including offering $0$.
{\em Actually} offering $o<0.5$ is certainly not optimal, but such a suboptimal offer can be easily
fixed by raising it to the best revenue point, i.e., to $\arg\max_{x\ge o}x(1-F_i(x))=0.5$. Another way to put it is that
the suboptimal offer can be fixed by pushing it up until the marginal profit becomes positive, i.e., to $\inf\{x|m_i(x)>0,x\ge o\}=0.5$.
The above example illustrates an advantage of marginal profit -- it cuts more slack to our learning procedure.
If we learn the allocation for this example by mimicking the greedy allocation based on either the virtual valuation
or the marginal profit, then
if we go with the virtual valuation,
the learned reserve must be {\em exactly} $0.5$. Any more or less is not optimal.
On the other hand, according to the marginal profit, $0.43$ is optimal, and so is $0.37$.
The exact optimal reserve can always be recovered by pushing up the offer via the revenue fix process. It should be noted that we {\em must} go through the trouble
of performing the revenue fix process if the learning target is based on the marginal profit. Otherwise, we run the risk of using a reserve of $0$, which is far from optimal.

\section{Training Parameters, Evaluation Details, and Hardware}
\label{sec:parameters}

The description below is for our main approach. For integration with RegretNet, please refer to Appendix~\ref{sec:rnet}.

\vspace{.1in}
\noindent {\bf Training:} We use an MLP with $2$ hidden layers to represent the allocation function. Each layer
contains $20$ nodes.  We use the Adam
optimizer with learning rate $0.001$. Unless otherwise specified, we allow the
optimizer to step $20,000$ times.  The batch size is $16$.  For counterexample-guided training, we use all counterexamples and the penalty term is multiplied by $1,000$.

The following description applies to all distributions except for {\sc worst10} and {\sc worst100} (Appendix~\ref{subsec:worst10and100}).
For $2$ bidders, we only need to train once to achieve a monotone allocation. No counterexample-guided training or monotonicity fix is needed.
This is true even for the $10$ adversarial
distributions generated using evolutionary computation (Appendix~\ref{subsec:ec}).
For $3$ to $5$ bidders, we train $10$ times as some trials failed to achieve monotonicity.
$10$ trials are more than enough. Even in the worst situation, at least $6$ of the $10$ trials ended up monotone without using counterexample-guided training or monotonicity fix.
For {\sc worst10} and {\sc worst100}, we do need counterexample-guided training and monotonicity fix. Otherwise,
near-optimal revenue cannot be achieved. The experiments on these
two distributions are presented in Appendix~\ref{sec:100trials} and \ref{subsec:worst10and100}.

\vspace{.1in}
\noindent {\bf Evaluation:} We evaluate an auction's revenue via Monte Carlo average with sample size $100,000$.
When monotonicity fix is needed, the sample size is reduced to $10,000$, as for each profile, monotonicity fix involves $n$ mixed-integer programs.

\noindent {\bf Hardware:} The only large-scale experiments are the experiments described in Figure~\ref{fig:bar}, Figure~\ref{fig:worst10} and Figure~\ref{fig:worst100}, where for each of the three selected distributions ($G5_6$, {\sc worst10}, {\sc worst100}), we performed $100$ training trails.
This is carried out on Adelaide University's high-performance cluster. Our performance bottleneck is the MIPs for monotonicity fix, which needs to be repeated $n\times 10,000$ times in evaluation, so our jobs are CPU intensive. (Running the auction once is always instant.) For each of the $300$ trials, we allocate one CPU core, 128 GB memory, and no GPU.
The CPU type is Intel(R) Xeon(R) Platinum 8360Y. Each trial takes from a few minutes (when monotonicity fix is not needed) to up to six hours.
We have to point
out that the wall clock time is highly inaccurate for the cluster we use (i.e., CPUs are shared; jobs may hang or slow down without notice).
We estimate that the total computation time is on the order of $500$ hours.

\section{Details of Baseline Auctions}
\label{sec:baseline}

{\sc Greedy}: The greedy allocation simply allocates the item to the bidder with
the highest marginal profit.  Based directly on the definition of marginal
profit~\cite{Papadimitriou11:On}, the greedy allocation's revenue can serve as the
revenue {\em upper bound}.  To calculate this upper bound, we use Monte Carlo simulation to calculate the expectation of the highest
marginal profit. For the vast majority of the distributions we use, the greedy
allocation is {\bf not} monotone, and therefore {\bf not} strategy-proof, which is
proved via counterexamples.
When counterexamples cannot be found, monotonicity/strategy-proofness is inconclusive.

{\sc AMD}: \cite{Papadimitriou11:On} proposed an FPTAS for deriving the optimal auction
for $2$ correlated bidders. The proposed algorithm first rounds down the bids to grid points, for example,
to multiples of $0.01$. The rest of the algorithm converts the instance to a maximum independent set instance on a bipartite graph.
We use AMD (Automated Mechanism Design~\cite{Conitzer02:Complexity}) as an alternative implementation.
We round the bids down to multiples of $0.01$. After this step, there are $100^2$ possible bid profiles.
For each bid profile, we create $2$ continuous variables to represent the payments and $2$ binary variables to represent
the allocation. We follow the standard AMD process. The expected revenue can be easily expressed as a linear expression of the variables. Strategy-proofness and individual rationality can also easily be expressed as linear inequalities involving the variables. The optimal auction for discrete bids is derived
via a mixed-integer program.
For $2$ bidders, the AMD solution takes only seconds. (This is completely expected. When the underlying problem has a polynomial-time solution, state-of-the-art MIP solvers often manage to solve it fast.) AMD does not scale to $3$ or more bidders. The FPTAS is also restricted to $2$ bidders. As mentioned in \cite{Papadimitriou11:On}, optimal auction design for correlated bidders becomes NP-hard to approximate when $n\ge 3$. For the above reasoning, AMD suffices as an alternative implementation.

{\sc Myerson}: We calculate every bidder's ironed virtual valuations based on their {\em marginal} distributions, instead
of their {\em conditional} distributions.
We then allocate the item to the bidder with the highest ironed virtual valuation. The item is not allocated if the
highest ironed virtual valuation is below $0$.
This auction guarantees monotonicity and therefore strategy-proofess, as all it does is to pretend that correlation
does not exist and runs the optimal auction for independent bidders.
The revenue of this auction is expected to be sub-optimal as it ignores correlation completely.

{\sc 2nd}: Second price auction with the optimal reserve price. The optimal reserve is calculated by simply trying
all multiples of $0.001$.
\newpage

\section{Hand-Crafted Grid Distributions Based on Magic Squares}
\label{sec:magic}

\begin{align*}
\textsc{saturn} &=
\begin{bmatrix}
4 & 9 & 2 \\ 3 & 5 & 7 \\ 8 & 1 & 6
\end{bmatrix}\\
\textsc{jupiter} &=
\begin{bmatrix}
4 & 14 & 15 & 1 \\ 9 & 7 & 6 & 12 \\ 5 & 11 & 10 & 8 \\ 16 & 2 & 3 & 13
\end{bmatrix}\\
\textsc{mars} &=
\begin{bmatrix}
11 & 24 & 7 & 20 & 3 \\ 4 & 12 & 25 & 8 & 16 \\ 17 & 5 & 13 & 21 & 9 \\ 10 & 18 & 1 & 14 & 22 \\ 23 & 6 & 19 & 2 & 15
\end{bmatrix}\\
\textsc{sol} &=
\begin{bmatrix}
6 & 32 & 3 & 34 & 35 & 1 \\ 7 & 11 & 27 & 28 & 8 & 30 \\ 19 & 14 & 16 & 15 & 23 & 24 \\ 18 & 20 & 22 & 21 & 17 & 13 \\ 25 & 29 & 10 & 9 & 26 & 12 \\ 36 & 5 & 33 & 4 & 2 & 31
\end{bmatrix}\\
\textsc{venus} &=
\begin{bmatrix}
22 & 47 & 16 & 41 & 10 & 35 & 4 \\ 5 & 23 & 48 & 17 & 42 & 11 & 29 \\ 30 & 6 & 24 & 49 & 18 & 36 & 12 \\ 13 & 31 & 7 & 25 & 43 & 19 & 37 \\ 38 & 14 & 32 & 1 & 26 & 44 & 20 \\ 21 & 39 & 8 & 33 & 2 & 27 & 45 \\ 46 & 15 & 40 & 9 & 34 & 3 & 28
\end{bmatrix}\\
\textsc{mercury} &=
\begin{bmatrix}
8 & 58 & 59 & 5 & 4 & 62 & 63 & 1 \\ 49 & 15 & 14 & 52 & 53 & 11 & 10 & 56 \\ 41 & 23 & 22 & 44 & 45 & 19 & 18 & 48 \\ 32 & 34 & 35 & 29 & 28 & 38 & 39 & 25 \\ 40 & 26 & 27 & 37 & 36 & 30 & 31 & 33 \\ 17 & 47 & 46 & 20 & 21 & 43 & 42 & 24 \\ 9 & 55 & 54 & 12 & 13 & 51 & 50 & 16 \\ 64 & 2 & 3 & 61 & 60 & 6 & 7 & 57
\end{bmatrix}\\
\textsc{luna} &=
\begin{bmatrix}
37 & 78 & 29 & 70 & 21 & 62 & 13 & 54 & 5 \\ 6 & 38 & 79 & 30 & 71 & 22 & 63 & 14 & 46 \\ 47 & 7 & 39 & 80 & 31 & 72 & 23 & 55 & 15 \\ 16 & 48 & 8 & 40 & 81 & 32 & 64 & 24 & 56 \\ 57 & 17 & 49 & 9 & 41 & 73 & 33 & 65 & 25 \\ 26 & 58 & 18 & 50 & 1 & 42 & 74 & 34 & 66 \\ 67 & 27 & 59 & 10 & 51 & 2 & 43 & 75 & 35 \\ 36 & 68 & 19 & 60 & 11 & 52 & 3 & 44 & 76 \\ 77 & 28 & 69 & 20 & 61 & 12 & 53 & 4 & 45
\end{bmatrix}
\end{align*}

The baseline auction {\sc Myerson} refers to Myerson's original optimal auction extended to correlated bidders, simply
ignoring correlation altogether. This is achieved by pretending that the distribution
is independent and using the marginal distribution of a bidder to calculate her ironed
virtual valuation.
To assess the amount of revenue lost by ignoring correlation, we resort to the {\em magic squares}.
These are 2D matrices whose every row and every column sums up to the same value.
Under grid distributions based on magic squares, every bidder's marginal distribution is always $U(0,1)$.
The magic squares listed in this section are from a 1531 book titled {\em De occulta philosophia}.
We experimented on these magic-square-based grid distributions.
The revenue gap between {\sc Myerson} (optimal if correlation does not exist) and our auction (proven near-optimal based on the upper bound) is the amount of revenue lost by ignoring correlation. For i.i.d. distributions, {\sc Myerson} is just the second
price auction with a reserve based on the i.i.d. distributions.

\section{Adversarial Distributions}
\label{sec:adversarial}

After experimenting on $47$ distributions, including $10$
randomly generated grid distributions for $n$ from $2$ to $5$, and $7$ hand-crafted grid distributions based on the magic squares in Appendix~\ref{sec:magic}, an
unexpected observation is that for all these distributions, a near-optimal
verifiably monotone allocation can be found without incorporating counterexample-guided training. The post-processing monotonicity fix step is therefore
also not needed.  For $2$ bidders, we always find a verifiably monotone
allocation with one trial.  For more bidders, we sometimes do end up with
non-monotone allocations, but this can be resolved by repeated trials.
We do not need an
excessively large number of trials.  We used $10$
trials in experiments. In the worst situation ($5$ bidders, distribution $G5_6$), we ended up with a verifiably
monotone allocation in $6$ out of $10$ trials.
That is, in expectation, we only need less
than $2$ trials to obtain a monotone allocation. Furthermore, for all these distributions, we achieve near-optimal revenue.

While the above observation shows the effectiveness of our approach,
a natural question to ask is whether these results are artifacts caused by our selection of distributions.
To test the limit of our approach, we experimented with two methods for generating adversarial distributions.
The first method is based on evolutionary computation. We managed to generate distributions under which the
greedy allocation is (probably) ``wrong'' for more bid profiles, in the sense that its decision is inconsistent with decisions obtained via Automated Mechanism Design. Nevertheless, our approach still delivers near-optimal results with one trial, without
any counterexample-guided training.
The evolutionary computation based
adversarial distributions did increase
the revenue gap between the greedy upper bound and the achieved revenue, from maximally $1.3\%$ for non-adversarial distributions to $2.1\%$.
The second method relies on a mixed-integer program to generate adversarial distributions under which
the greedy allocation maximally ``overestimates'' the revenue. This method successfully leads to two adversarial
distributions for which we do need counterexample-guided training and monotonicity fix.
That is, simply relying on luck (i.e., train again until we reach a verifiably monotone allocation with near-optimal
revenue) no longer works. We do need the full suite of techniques (verification, counterexample-guided training and
post-processing monotonicity fix) to achieve near-optimal revenue.

\subsection{Adversarial Distributions via Evolutionary Computation}
\label{subsec:ec}

Our approach aims to find a monotone interpolation of the greedy allocation. One adversarial situation is when the learning
target, i.e., the greedy allocation, makes allocation mistakes for many bid profiles.
In this section, we apply evolutionary computation to search for a distribution under which the greedy allocation
differs the most from the ``correct'' allocation (i.e., the optimal allocation).

However, the correct allocation is unknown. We settle for the AMD allocation instead, which we expect to be mostly
correct.
The main gist of AMD is that we can round the bids down to multiples of $1/H$.
Afterwards, the optimal auction for these discrete bids can be derived via mixed-integer programming.
Since AMD is only scalable for $2$ bidders, we focus on $2$ bidders.
Going through the $H^2$ bid profiles, we count the number of times greedy and AMD are
inconsistent. This counter is used as the fitness function for evolutionary computation -- we want its value to go up.
In our experiment, we set $H$ to $20$ to speed up the fitness evaluation.

We apply the simple (1+1)EA algorithm.
We start with the grid distribution $G5_s$ for $s$ from $0$ to $9$.
We use $d$ to denote the current distribution.
In every evolutionary round, for every density value in the 2D matrix representing $d$, we replace the original density value
$x$ by $0.9\cdot x + u$, where $u$ is drawn from $U[0,1]$. After mutation, the density values are normalized to ensure that the total probability
equals $1$.
We evaluate the fitness of the mutated distribution $d'$. If it is better (i.e., more adversarial), then we keep it by replacing
$d$ by $d'$. Otherwise, we
throw $d'$ away and go back to $d$. We do the above for $100$ rounds. The resulting distribution is recorded as $EA(G5_s)$.

Despite the effort, under the most adversarial distribution we found via evolutionary computation, greedy and AMD only differ under $3.5\%$ of the bid profiles. This empirically suggests that the greedy allocation is a high quality learning target even
for (somewhat) adversarial distributions, justifying our approach. For the evolved adversarial distributions, our neural network interpolation approach still produces near-optimal revenue as shown in Table~\ref{tab:n2gan}.
Among all synthetic, hand-crafted and adversarial distributions we experimented on, $EA(G5_2)$ leads to the worst revenue gap between the greedy upper bound
and the achieved revenue, which is $2.1\%$. (For non-adversarial distributions, the maximum revenue gap is $1.3\%$.)

\begin{table*}[h]
\centering
\caption{$2$ bidders, evolutionary computation generated adversarial distributions}
\label{tab:n2gan}
\begin{footnotesize}
\begin{tabular}{lll|lllllll}
\toprule
{\sc Distrib.} &  {\sc Greedy} & {\sc AMD} & {\sc Myer.} & {\sc 2nd} & {\sc RNet} & {\sc MP} \\
\midrule

$EA(G5_0)$ & $0.3935$ & $0.3918$ & $0.3857$ & $0.3869$ & $0.3875$ & $\textbf{0.3905}$\\
& $\pm0.001$ & & $\pm0.0007$ & $\pm0.0008$ & $\pm0.0008$ & $\pm0.0008$\\
$EA(G5_1)$ & $0.4143$ & $0.4111$ & $0.4030$ & $0.4010$ & $0.4028$ & $\textbf{0.4106}$\\
& $\pm0.001$ & & $\pm0.0008$ & $\pm0.0008$ & $\pm0.0008$ & $\pm0.0008$\\
$EA(G5_2)$ & $0.4311$ & $0.4279$ & $0.4137$ & $0.4116$ & $0.3836$ & $\textbf{0.4220}$\\
& $\pm0.001$ & & $\pm0.0007$ & $\pm0.0007$ & $\pm0.001$ & $\pm0.0008$\\
$EA(G5_3)$ & $0.4165$ & $0.4140$ & $0.3999$ & $0.3986$ & $0.4033$ & $\textbf{0.4117}$\\
& $\pm0.001$ & & $\pm0.0008$ & $\pm0.0008$ & $\pm0.0008$ & $\pm0.0008$\\
$EA(G5_4)$ & $0.4271$ & $0.4228$ & $0.4152$ & $0.4156$ & $0.4175$ & $\textbf{0.4260}^*$\\
& $\pm0.001$ & & $\pm0.0007$ & $\pm0.0008$ & $\pm0.0008$ & $\pm0.0008$\\
$EA(G5_5)$ & $0.4374$ & $0.4320$ & $0.4277$ & $0.4246$ & $0.4258$ & $\textbf{0.4320}^*$\\
& $\pm0.001$ & & $\pm0.0007$ & $\pm0.0008$ & $\pm0.0008$ & $\pm0.0007$\\
$EA(G5_6)$ & $0.4381$ & $0.4362$ & $0.4256$ & $0.4279$ & $0.4307$ & $\textbf{0.4336}$\\
& $\pm0.001$ & & $\pm0.0008$ & $\pm0.0008$ & $\pm0.0008$ & $\pm0.0008$\\
$EA(G5_7)$ & $0.4025$ & $0.3998$ & $0.3912$ & $0.3934$ & $0.3958$ & $\textbf{0.4002}^*$\\
& $\pm0.001$ & & $\pm0.0007$ & $\pm0.0008$ & $\pm0.0008$ & $\pm0.0008$\\
$EA(G5_8)$ & $0.4355$ & $0.4320$ & $0.4211$ & $0.4224$ & $0.4199$ & $\textbf{0.4324}^*$\\
& $\pm0.001$ & & $\pm0.0008$ & $\pm0.0008$ & $\pm0.0008$ & $\pm0.0008$\\
$EA(G5_9)$ & $0.4141$ & $0.4107$ & $0.4068$ & $0.4076$ & $0.4051$ & $\textbf{0.4097}$\\
& $\pm0.001$ & & $\pm0.0008$ & $\pm0.0008$ & $\pm0.0008$ & $\pm0.0009$\\

\bottomrule
\end{tabular}
\end{footnotesize}
\end{table*}

\subsection{Adversarial Distributions via Mixed-Integer-Programming}
\label{subsec:worst10and100}

We explore another heuristic direction for finding adversarial distributions.
We use $R^G$ to denote the greedy revenue, which is generally not attainable as greedy is often not strategy-proof.
$R^G$ serves as the revenue upper bound. We use $R^*$ to denote the optimal revenue.
We aim to find the distribution that maximizes the ratio $\frac{R^G}{R^*}$.
That is, we are searching for distributions under which the greedy allocation significantly overestimates the revenue,
which would make greedy an unsuitable learning target.

We still work within the structure of the grid distributions, focusing on $5\times 5$ matrices as illustrated below.
Our goal is to search for a set of $25$ values for the $p_{i,j}$, so that the corresponding grid distribution is adversarial based on the metric mentioned above.
\begin{align*}
    \textsc{worst}(\lambda) &=
\begin{bmatrix}
p_{0,4} & p_{1,4} & p_{2,4} & p_{3,4} & p_{4,4}
\\
p_{0,3} & p_{1,3} & p_{2,3} & p_{3,3} & p_{4,3}
\\
p_{0,2} & p_{1,2} & p_{2,2} & p_{3,2} & p_{4,2}
\\
p_{0,1} & p_{1,1} & p_{2,1} & p_{3,1} & p_{4,1}
\\
p_{0,0} & p_{1,0} & p_{2,0} & p_{3,0} & p_{4,0}
\end{bmatrix}\\
\end{align*}
We construct a mixed-integer program to find these $25$ values.
There are $25$ continuous variables, i.e., the $p_{i,j}$, ranging from $1$ to $\lambda$, where $\lambda$ is a distribution parameter.
We derived two distributions called {\sc worst10} (setting $\lambda=10$) and {\sc worst100} (setting $\lambda=100$).

We face two challenges when constructing the mixed-integer program:

The first challenge is that the marginal profit is not linear in the probability densities. To make it
linear, we re-interpret the above matrix as a discrete distribution. Every
bidder's bid can only be from the following $5$ values:
$\{0,0.2,0.4,0.6,0.8\}$.  $p_{0,0}$ refers to the probability that both bidders
bid $0$s.  $p_{1,3}$ refers to the probability that bidder $1$ bids $0.2$ and
bidder $2$ bids $0.6$.  Essentially, for the original 2D grid distribution,
there are $25$ uniform sub-squares, we assign all probability mass to the
bottom left corner of the sub-square to make the distribution discrete.  By
pretending that the distribution is discrete, the marginal profits become
linear in the $p_{i,j}$ with the help of axillary binary variables.
We use $r_{i,j}^k$ with $k \in \{1,2\}$ and $0\le i,j\le 4$ to represent the maximum revenue we
can extract from bidder $k$ when bidder $1$ bids $i/5$ and bidder $2$ bids $j/5$.
We have
\[r_{i,j}^1 = \max_{i'\ge i}\left(\frac{i'}{5}\sum_{i''\ge i'}p_{i'',j}\right)\]
\[r_{i,j}^2 = \max_{j'\ge j}\left(\frac{j'}{5}\sum_{j''\ge j'}p_{i,j''}\right)\]
The marginal profit of bidder $k$ when bidder $1$ bids $i/5$ and bidder $2$ bids $j/5$ is denoted as $m_{i,j}^k$, which equals the following. (This is not the ``per density'' version. Please refer to Appendix~\ref{sec:appendixvv}.)
\[m_{i,j}^1 = \max\{r_{i,j}^1-r_{i+1,j}^1,0\}\]
\[m_{i,j}^2 = \max\{r_{i,j}^2-r_{i,j+1}^2,0\}\]
The greedy revenue $R^G$ is then
\[\sum_{0\le i,j\le 4} \max\{m_{i,j}^1,m_{i,j}^2\}\]
The second challenge is that the optimal revenue is unknown.
We settle for an alternative revenue. We define two auctions for
discrete bids.
Auction $1$ allocates the item to bidder $1$ if and only if $b_1\ge b_2$.
Auction $2$ allocates the item to bidder $1$ if and only if $b_1>b_2$.
The only difference between these two auctions is the way they perform tie-breaking. Note that tie-breaking is consequential for discrete distributions.
We evaluate the revenue of these two auctions using marginal profit.
The revenue under auction $1$ is
\[\sum_{0\le j\le i\le 4}m_{i,j}^1 + \sum_{0\le i<j\le 4}m_{i,j}^2\]
The revenue under auction $2$ is
\[\sum_{0\le j<i\le 4}m_{i,j}^1 + \sum_{0\le i\le j\le 4}m_{i,j}^2\]
The maximum revenue between these two auctions is denoted as $R^S$.

Since the revenue of $R^S$ is evaluated in terms of marginal profit, the revenue fix is already automatically included. The revenue fix basically adds a reserve for each bidder and the reserve can depend on the other bid. The optimal reserve is already reflected in the
marginal profits.

So far, all values mentioned above can be expressed as linear expressions of the $p_{i,j}$ (with the help of many
auxiliary binary variables to represent ``max'').
We finally add a constraint $R^G\ge \alpha \cdot R^S$, where $\alpha$ is a constant. We search for the largest constant $\alpha$ that still makes the above inequality holds, which can be solved via a mixed-integer program (i.e., via feasibility check). The $p_{i,j}$ values corresponding to the largest $\alpha$ characterize the adversarial distribution.
Below are the solutions for $\lambda\in\{10,100\}$:
\begin{align*}
    \textsc{worst10} &=
\begin{bmatrix}
1 & 1 & 1 & 1 & 1 \\ 1 & 10 & 9 & 1 & 10 \\ 1 & 1 & 1 & 1 & 1 \\ 1 & 1 & 1 & 2.59 & 1 \\ 1 & 1 & 1 & 7.77 & 1
\end{bmatrix}\\
    \textsc{worst100} &=
\begin{bmatrix}
1 & 1 & 1 & 1 & 1 \\ 1 & 100 & 1 & 1 & 1 \\ 1 & 1 & 48.5 & 24.25 & 72.75 \\ 1 & 14.25 & 1 & 1 & 3 \\ 1 & 42.75 & 1 & 1 & 1
\end{bmatrix}
\end{align*}
For {\sc worst10} and {\sc worst100}, our approach still manages to produce near-optimal
revenue, as shown in Table~\ref{tab:n2gan2}. However, these two distributions indeed are significantly more challenging to handle. Unlike other distributions
where we can completely skip counterexample-guided training and monotonicity fix (by resorting to repeated trials), for {\sc worst10} and {\sc worst100},
we do need the full suite of techniques. More detailed case studies on these two distributions are presented in Appendix~\ref{sec:100trials}.

\begin{table*}[h]
\centering
\caption{$2$ bidders, mixed-integer-programming generated adversarial distributions}
\label{tab:n2gan2}
\begin{footnotesize}
\begin{tabular}{lll|lllllll}
\toprule
    {\sc Distrib.} &  {\sc Greedy} & {\sc AMD} & {\sc Myer.} & {\sc 2nd} & {\sc RNet} & {\sc MP} \\
\midrule

    {\sc worst10} & $0.4401$ & $0.4379$ & $0.3733$ & $0.3919$ & $0.4130$ & $\textbf{0.4441}^*$\\
                  & $\pm0.001$ & & $\pm0.0009$ & $\pm0.0006$ & $\pm0.0007$ & $\pm0.003$\\

    {\sc worst100} & $0.4678$ & $0.4565$ & $0.3885$ & $0.3805$ & $0.4126$ & $\textbf{0.4632}^*$\\
                  & $\pm0.001$ & & $\pm0.0007$ & $\pm0.0004$ & $\pm0.0006$ & $\pm0.003$\\

\bottomrule
\end{tabular}
\end{footnotesize}
\end{table*}

\section{Case Study on the Benefit of Counterexample-Guided Training and Monotonicity Fix}
\label{sec:100trials}

To test the necessity and effectiveness of counterexample-guided training and monotonicity fix, we run the following
experiment. For each selected distribution, we run $100$ training trials to reach statistically meaningful conclusions.
Each training trial proceeds as follows:

\begin{itemize}

    \item Train $20,000$ (optimizer) steps without counterexample-guided training. We call the result at the
        end of $20,000$ steps the {\em stage-one} result.

    \item We verify whether the stage-one allocation is monotone and evaluate its revenue.
        If the stage-one allocation is already monotone, then we do not need to apply monotonicity fix when evaluating its revenue.
        Otherwise, the monotonicity fix is used.

    \item We continue to train another $10,000$ steps. If the stage-one allocation is not monotone, then we switch
        to counterexample-guided training and use the monotonicity violations from stage-one's evaluation as counterexamples.
        If the stage-one allocation is already monotone, then we do not need to switch to counterexample-guided training (and we do not have counterexamples anyway).
        We call the end result the {\em stage-two} result.

    \item We verify whether the stage-two allocation is monotone and evaluate its revenue.
        Similar to the situation for stage-one, we apply monotonicity fix when necessary.

\end{itemize}

\begin{figure*}[h!]
  \centering
  \includegraphics[width=0.9\textwidth]{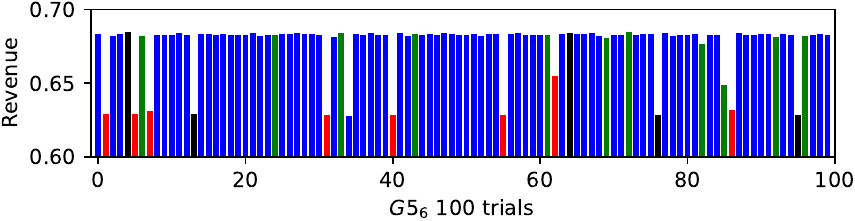}
    \caption{
    While green and black reached slightly higher highs, the gap between blue and green/black is statistically insignificant
    considering the standard errors.
    In conclusion, counterexample-guided training and monotonicity fix are not needed for this distribution.}
  \label{fig:bar}
\end{figure*}

\begin{figure*}[h!]
  \centering
  \includegraphics[width=0.9\textwidth]{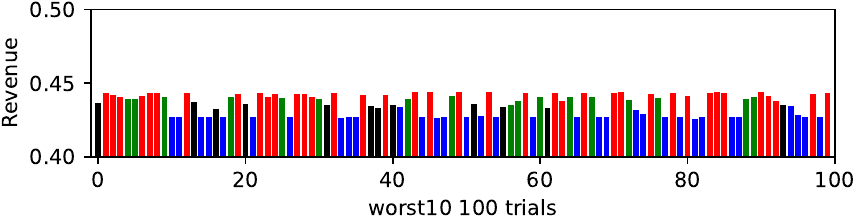}
    \caption{Red and green reached higher, so counterexample-guided training and monotonicity fix do help.
    We do need the whole suite of techniques to reach near-optimal revenue.}
  \label{fig:worst10}
\end{figure*}

\begin{figure*}[h!]
  \centering
  \includegraphics[width=0.9\textwidth]{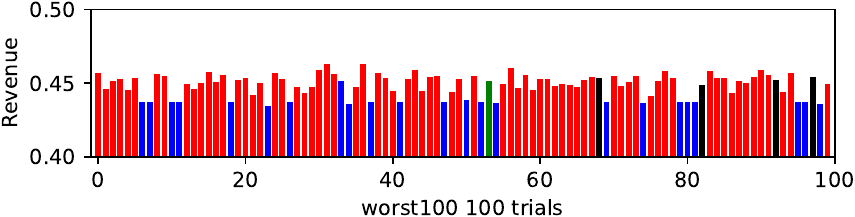}
    \caption{Red reached higher, so counterexample-guided training and monotonicity fix do help.
    We do need the whole suite of techniques to reach near-optimal revenue.}
  \label{fig:worst100}
\end{figure*}

The trials were divided into four categories, which are represented using different colours in Figure~\ref{fig:bar}, Figure~\ref{fig:worst10} and Figure~\ref{fig:worst100}:

\begin{itemize}

    \item Blue: Stage-one and two are both monotone. That is, blue results do not involve counterexample-guided training
        or monotonicity fix.

    \item Red: Stage-one and two are both not monotone. Counterexample-guided training and monotonicity fix are both used.

    \item Green: Stage-one is not monotone and stage-two is monotone. That is, counterexample-guided training managed to guide the
        allocation to monotonicity. Both counterexample-guided training and monotonicity fix are used.

    \item Black: Stage-one is monotone but stage-two is not monotone. That is, with additional training, the network learned
        more details and turned not monotone. Monotonicity fix is used but not counterexample-guided training.

\end{itemize}

We perform a detailed analysis of the following $3$ distributions. As mentioned earlier, in this experiment, we run $100$ trails for each distribution. For $3$ distributions,
we estimate that the total computation time is on the order of $500$ hours, so we cannot afford to test every distribution.

\begin{itemize}

    \item $G5_6$ for $5$ bidders: This distribution is selected because it is more difficult
        to reach monotonicity under $G5_6$ compared to the other randomly generated distributions.
        When constructing Table~\ref{tab:n3-1}, Table~\ref{tab:n4-1} and Table~\ref{tab:n5-1}, for every distribution we tried $10$ times and recorded how many times the result is monotone. For $G5_6$, $6$ out of $10$ times are monotone. $6/10$ appears to
        be a fairly high success rate, but is actually the worst among the randomly generated distributions.

    \item {\sc worst10} and {\sc worst100} for $2$ bidders: These are adversarial distributions designed to test the limit of our approach. They are derived in Appendix~\ref{subsec:worst10and100}.

\end{itemize}

The experimental results are summarized in Table~\ref{tab:100trials}.
The individual distribution's reports are given in Figure~\ref{fig:bar}, Figure~\ref{fig:worst10} and Figure~\ref{fig:worst100}.

\begin{table*}[h]
    \caption{Summary of best revenue per category, where $100$ trials are divided into $4$ categories based on the monotonicity status at the end of stage-one and two: $M$=Monotonicity; $\overline{M}$=non-monotonicity; $M\rightarrow \overline{M}$ means the stage-one result is monotone and the stage-two result is not monotone}
  \label{tab:100trials}
  \centering
    \begin{footnotesize}
  \begin{tabular}{lllll}
    \toprule
      {\sc Distrib.} &  {\sc Blue} & {\sc Red} & {\sc Green} & {\sc Black} \\
       &  $M\rightarrow M$ & $\overline{M}\rightarrow \overline{M}$ & $\overline{M}\rightarrow M$ & $M\rightarrow \overline{M}$ \\
    \midrule

      $G5_6$ ($n=5$) & $0.6841\pm 0.0006$ & $0.6551\pm 0.002$ & $0.6847\pm 0.002$ & $0.6844\pm 0.002$ \\

      {\sc worst10} ($n=2$)  &  $0.4347\pm 0.0009$  & $0.4441\pm 0.003$ & $0.4410\pm 0.003$ & $0.4370\pm 0.003$ \\

      {\sc worst100} ($n=2$)  & $0.4517\pm 0.0008$ & $0.4632\pm 0.003$ & $0.4516\pm 0.002$ & $0.4544\pm 0.0008$ \\
    \bottomrule
  \end{tabular}
    \end{footnotesize}
\end{table*}

\section{Details of RegretNet Implementation}
\label{sec:rnet}

RegretNet \cite{Dutting19:Optimal} implements two neural networks, one for allocation and
one for payments. In our RegretNet implementation, the two networks have the following
properties:

\begin{itemize}
    \item Each network takes the $n$ bids as inputs.
    \item Each network is a multilayer perceptron (MLP) with ReLU activation. In our implementation, both networks have $2$ intermediate layers each with width $50$. This is larger than the networks used in the main approach, which is $2$ layers with width $20$.
    \item The allocation network has $n+1$ outputs where the first $n$ outputs correspond to the $n$ bidders and the last output corresponds to no allocation. We take the argmax among the $n+1$ outputs to decide which bidder wins. In training, we take softmax on the outputs.
    \item The payment network has $n$ outputs where the $i$-th output decides bidder $i$'s payment if the item is allocated to bidder $i$.
\end{itemize}

RegretNet is unsupervised. The training samples are bid profiles drawn according to the given correlated distribution.
Unlike the main approach, where we need the distribution, as otherwise we cannot calculate the marginal profits, RegretNet does not need the access to the distribution. Samples are all it needs.

Following RegretNet's notation, we define the set of bidders as $N$, denote the model parameters as $w$, denote $v_i$ as the true valuation of bidder $i$.
Bidder $i$'s utility is $u_i^w(v_i;(v_i',v_{-i}))$ when she has valuation $v_i$ and bids $v_i'$.
Bidder $i$'s payment is denoted as $p_i^w(v)$.

The aim of RegretNet is to maximize the expected total payment $\mathbb{E}[\sum_{i\in N}p_i^w(v)]$ while ensuring strategy-proofness as well as ex post individual rationality. During training, we use a batch of $L$ bid profiles to get an unbiased estimation $\frac 1 L \sum_{l=1}^{L}\sum_{i\in N}p_i^{w,l}(v)$.

The strategy-proofness constraint is given by minimizing the ``Regret'' defined as the sum of how profitable the optimal deviation is for every bidder. Formally,
\[rgt_i(w)=\mathbb{E}\left[\max\limits_{v_i'\in V_i}u_i^w(v_i;(v_i',v_{-i}))-u_i^w(v_i;(v_i,v_{-i}))\right].\]
During training, we use a batch of bid profiles to give an unbiased estimation $\widehat{rgt}_i(w)$.

Individual rationality is ensured by taking the sigmoid of the payment network's outputs, so that the range becomes $[0,1]$, then times the bid. This method can ensure that the payment from each bidder is nonnegative and no more than the bid itself, thus satisfying individual rationality.

The loss function used in unsupervised training is as follows:
\[Loss(w;\lambda)=-\frac 1 L \sum_{l=1}^{L}\sum_{i\in N}p_i^{w,l}(v)+\sum_{i\in N}\lambda_i \widehat{rgt}_i(w). \]
In the original implementation of RegretNet \cite{Dutting19:Optimal}, there is also a second order term, which
we dropped. %
The factor $\lambda_i$'s are initialized to 50 and updated periodically. We set the update interval to be 20 backward steps. The original RegretNet uses a gradient descent method to find the optimal update direction and amount, while in our implementation, we set it higher if $\widehat{rgt}_i(w)$ is larger than $0.0001$, and smaller if it is less than it. $0.0001$ is the largest regret that we tolerate.

Since the allocation network is using the same network architecture as the main approach, i.e., MLP+ReLU, the existing suite of MIP-based techniques
still apply, including verification, counterexample-guided training, and monotonicity fix.
After training, the payment network is thrown away and we only evaluate the allocation network.
The existing evaluation process from the main approach still applies.
For all results on RegretNet listed in this paper, the auction is verified $100\%$ strategy-proof (no counterexample-guided training was needed).
For the distributions in Table~\ref{tab:n2-1}, our RegretNet implementation is able to outperform {\sc Myerson} and {\sc 2nd},
and achieve revenue that is reasonably close to our main approach, noting again that our main approach requires
more than just the samples, but also the exact distribution.
RegretNet also outperforms {\sc Myerson} and {\sc 2nd} for $8$ out of $12$ adversarial distributions (Table~\ref{tab:n2gan}
and Table~\ref{tab:n2gan2}).
On the other hand, for the distributions in Table~\ref{tab:n3-1}, Table~\ref{tab:n4-1} and Table~\ref{tab:n5-1} (i.e., $n=3,4,5$),
the revenue achieved is less ideal. We set the same training budget for our main approach and RegretNet, which is $20,000$
optimizer steps. For $3$ or more bidders, perhaps RegretNet requires a lot more training resources. After all, it trains
two networks instead of one and the networks are significantly larger.
We did not perform excessive hyper-parameter tuning for the RegretNet approach as RegretNet is not the main
focus of this paper (mostly used as a baseline). Furthermore, given its larger size, it is a lot more time consuming to work with. For example,
for the $10$ auctions in Table~\ref{tab:n5-1}, MIP-based verification takes $21$ to $100$ minutes.

\section{Alternative Network Architectures}
\label{sec:altnet}

The concept of monotonicity represented by the min-max networks~\cite{Sill97:Monotonic} is different from the concept of allocation monotonicity in our mechanism design context. For example, in our model, the $i$-th input of the allocation network is agent $i$'s bid and the $i$-th output of the allocation network represents the priority value of agent $i$. Whoever has the highest priority value wins. Allocation monotonicity basically requires the following: if the $i$-th output is currently the highest, then when we increase the $i$-th input, we want the $i$-th output to still be the highest. The $i$-th output does not have to be monotone with respect to the $i$-th input -- it is fine that an agent's priority value drops when she increases her bid, as long as she still wins (for example, the other agents' priority values may drop more, or the winner's priority value drop is not big enough to go below another agent). It is not clear how to reconcile the difference between monotonicity in terms of values between inputs and outputs and allocation monotonicity. In other words, it is not clear how to use the min-max networks to represent the whole space of strategy-proof allocations.

There are two ways to use min-max networks to represent allocation functions in our setting.
The first is MyersonNet~\cite{Dutting19:Optimal}. Under MyersonNet, agent $i$'s bid $b_i$ is mapped to $i$'s allocation priority value $f_i(b_i)$, where $f_i$ is nondecreasing
function represented by a min-max network. Whoever has the highest priority value wins. This representation is strategy-proof, but the following
example shows that MyersonNet cannot represent the full space of strategy-proof mechanisms and sometimes it leads to significant revenue loss.
In our experiments, we used the fully strategy-proof variant of RegretNet~\cite{Dutting19:Optimal} as a baseline, which is more
expressive than MyersonNet.

{\em Example:} We have two bidders. Bidder $2$'s value is either $0$ or $\epsilon$ (an infinitesimal value), {\em i.e.}, almost all revenue will come from bidder $1$. When bidder $2$'s value is $0$, bidder $1$'s conditional distribution is $D_0$. When bidder $2$'s value is $\epsilon$, bidder $1$'s conditional distribution is $D_1$. That is, bidder $2$ serves the purpose of sending a signal that tells us whether bidder $1$'s
distribution is $D_0$ or $D_1$.
Since almost all revenue comes from bidder $1$, a near-optimal auction has the following form: When bidder $2$'s value is $0$, the auction becomes a take-it-or-leave-it auction for bidder $1$ with an optimal reserve price derived based on $D_0$, which we call $\textsc{Reserve}(D_0)$. When bidder $2$'s value is $\epsilon$, the auction becomes a take-it-or-leave-it auction for bidder $1$ with an optimal reserve price derived on $D_1$, which we call $\textsc{Reserve}(D_1)$. We further assume that $\textsc{Reserve}(D_0) > \textsc{Reserve}(D_1)$. MyersonNet cannot express the above auction. Under MyersonNet, when bidder $2$'s bid increases, bidder $1$'s winning price never drops.
Basically, under MyersonNet, bidder $1$ must face sub-optimal reserve under either $D_0$ or $D_1$ (or both), and it is easy to
construct distributions where sub-optimal reserves significantly impact the revenue.

\cite{Roughgarden13:Optimal} mentioned the {\em Lopomo assumption}. Using our terminology, if a bidder's allocation priority value is increasing in her own bid and decreasing in others' bids, then allocation monotonicity is easily satisfied. We can certainly use the min-max networks to model such kind of allocations, but the auction in the above example cannot be expressed. In the above example near-optimal auction, bidder $1$, without changing her own bid, can go from losing to winning when bidder $2$ increases her bid, which goes against
the Lopomo assumption.

We further present experimental results on four additional approaches:

\begin{itemize}

    \item {\sc MyerNet}~\cite{Dutting19:Optimal}: Agent $i$'s bid $b_i$ is mapped
        to $f_i(b_i)$, where $f_i$ is a nondecreasing function.
        $f_i(b_i)$ represents agent $i$'s allocation priority value.
        The agent with the highest priority value wins the item.
        If all agents' priority values are below $0$, then the
        item is thrown away.
        The monotonicity of the $f_i$ guarantees strategy-proofness.
        The $f_i$ are represented using the
        monotone min-max networks~\cite{Sill97:Monotonic}.
        We train the $f_i$ using the standard RegretNet~\cite{Dutting19:Optimal} approach.
        We simultaneously train $n+1$ networks.
        Each $f_i$ is a min-max network with $5$ ``max'' groups and each group contains $5$ nodes.
        There is also a shared payment network, which is a MLP
        with $2$ hidden layers and $20$ nodes for each layer.
        (Despite the tiny networks, in practise, it is already fairly expensive to train, with more details given below.)
        The payment network is only used during training.
        After training, the payment network is thrown away and the ``correct'' payments are derived from the (architecturally guaranteed) monotonic allocation function.
        All other implementation details follow our RegretNet implementation described in Appendix~\ref{sec:rnet}.

    \item {\sc MinMax}: This is a generalised version of
        {\sc MyerNet}. Agent $i$'s priority value $f_i(b_i, b_{-i})$ now also depends on the others' bids. $f_i$ is nondecreasing in $b_i$ and nonincreasing
        in every dimension in $b_{-i}$.
        This representation also guarantees strategy-proofness.
        We also use the min-max networks to represent
        the $f_i$.\footnote{By flipping the sign of input
        dimension $i$, we can change the output from being nondecreasing to nonincreasing in dimension $i$.}
        Same as {\sc MyerNet}, each $f_i$ is a min-max network with $5$ ``max'' groups and each group contains $5$ nodes.
        The only difference is that the input dimension of
        $f_i$ is now $n$ instead of $1$.
        All other settings are the same as {\sc MyerNet}.

\end{itemize}

{\em On training difficulty:} Both {\sc MyerNet} and {\sc MinMax}
are fairly expensive to train. For experiments reported in this subsection, every instance takes about 12 hours on average to train.
We conducted our experiments ($20$ instances) in parallel on a high-performance cluster with Intel 8360Y CPUs. As mentioned earlier, in our
experiments, every min-max network contains only $5$ groups and each group contains $5$ nodes. If we enlarge the network size to $10$ groups and $10$
nodes each group, then training takes substantially longer.
Based on our estimation, {\em each} instance takes $30+$ hours with an Nvidia A100 GPU.

Below we present two modified approaches that are significantly more
scalable, by adopting the neural network interpolation idea
from our paper.

\begin{itemize}

    \item {\sc MyerNet+}: We apply our paper's main approach
        to improve {\sc MyerNet}. That is, we apply supervised
        training to train only the allocation function (i.e.,
        the $f_i$). The supervision goal is to train
        the $f_i$ to replicate exactly
        Myerson's greedy allocation -- the agent with the highest
        (conditional) virtual valuation should win the item and the item is thrown
        away if all virtual valuations are below $0$.

        Each $f_i$ is represented using a min-max network
        with $10$ ``max'' groups and each group contains $10$
        nodes. All other implementation details follow
        our main approach described in Appendix~\ref{sec:parameters}.
        Each training instance takes around $35$ minutes with an Intel 8360Y CPU.

        There are several reasons why {\sc MyerNet+} is significantly faster: 1) there is no need to calculate regret, which is
        the most time consuming step; 2) we switch from unsupervised
        training to supervised; 3) we have one less network to
        train as the payment network is no longer needed.

    \item {\sc MinMax+}: Same as {\sc MyerNet+}, we apply
        our paper's main approach to improve {\sc MinMax}.
        Same as the case for {\sc MyerNet+}, the supervision goal is to train the $f_i$ to replicate exactly Myerson's greedy allocation.
        Each $f_i$ is also represented using a min-max network with $10$ ``max'' groups and each group contains $10$
        nodes.  Each training instance also takes around $35$ minutes with an Intel 8360Y CPU.

\end{itemize}

\begin{table*}[h!]
\centering
\caption{Additional experiments for $5$ bidders\\
    $\textsc{Gap}(\textsc{MP})=0.87\%$,$\textsc{Gap}(\textsc{Myer.})=2.4\%$ (note: $\textsc{Gap}(\textsc{2nd})$ is also $2.4\%$)\\
    $\textsc{Gap}(\textsc{MyerNet+})=3.3\%$,$\textsc{Gap}(\textsc{MyerNet})=19\%$\\
    $\textsc{Gap}(\textsc{MinMax+})=3.4\%$,$\textsc{Gap}(\textsc{MinMax})=18\%$}
\label{tab:iclr}
\begin{footnotesize}
\begin{tabular}{ll|llllll}
\toprule
{\sc Distrib.} &  {\sc Greedy} & {\sc MyerNet} & {\sc MinMax} & {\sc MyerNet+} & {\sc MinMax+} & {\sc Myer.} & {\sc MP} \\
\midrule

$G5_0$ & $0.6862$ & $0.5754$ & $0.5135$ & $0.6630$ & $0.6625$ & $0.6695$ & $\textbf{0.6799}$\\
& $\pm0.0008$ & $\pm0.0005$ & $\pm0.0008$ & $\pm0.0006$ & $\pm0.0006$ & $\pm0.0006$ & $\pm0.0006$\\
$G5_1$ & $0.6901$ & $0.5746$ & $0.4503$ & $0.6681$ & $0.6670$ & $0.6737$ & $\textbf{0.6846}$\\
& $\pm0.0008$ & $\pm0.0005$ & $\pm0.0007$ & $\pm0.0006$ & $\pm0.0006$ & $\pm0.0006$ & $\pm0.0006$\\
$G5_2$ & $0.6838$ & $0.5216$ & $0.5688$ & $0.6642$ & $0.6635$ & $0.6709$ & $\textbf{0.6783}$\\
& $\pm0.0008$ & $\pm0.0006$ & $\pm0.0006$ & $\pm0.0006$ & $\pm0.0006$ & $\pm0.0006$ & $\pm0.0006$\\
$G5_3$ & $0.6894$ & $0.6549$ & $0.4469$ & $0.6665$ & $0.6659$ & $0.6724$ & $\textbf{0.6838}$\\
& $\pm0.0008$ & $\pm0.0006$ & $\pm0.0009$ & $\pm0.0006$ & $\pm0.0006$ & $\pm0.0006$ & $\pm0.0006$\\
$G5_4$ & $0.6853$ & $0.5809$ & $0.6270$ & $0.6632$ & $0.6627$ & $0.6692$ & $\textbf{0.6787}$\\
& $\pm0.0008$ & $\pm0.0006$ & $\pm0.0005$ & $\pm0.0006$ & $\pm0.0006$ & $\pm0.0006$ & $\pm0.0006$\\
$G5_5$ & $0.6850$ & $0.6051$ & $0.5904$ & $0.6650$ & $0.6642$ & $0.6717$ & $\textbf{0.6809}$\\
& $\pm0.0008$ & $\pm0.0005$ & $\pm0.0006$ & $\pm0.0006$ & $\pm0.0006$ & $\pm0.0006$ & $\pm0.0006$\\
$G5_6$ & $0.6902$ & $0.6139$ & $0.6369$ & $0.6661$ & $0.6658$ & $0.6710$ & $\textbf{0.6839}$\\
& $\pm0.0008$ & $\pm0.0006$ & $\pm0.0005$ & $\pm0.0006$ & $\pm0.0006$ & $\pm0.0006$ & $\pm0.0006$\\
$G5_7$ & $0.6900$ & $0.6041$ & $0.6161$ & $0.6659$ & $0.6655$ & $0.6719$ & $\textbf{0.6824}$\\
& $\pm0.0008$ & $\pm0.0006$ & $\pm0.0006$ & $\pm0.0006$ & $\pm0.0006$ & $\pm0.0006$ & $\pm0.0006$\\
$G5_8$ & $0.6868$ & $0.2196$ & $0.6210$ & $0.6644$ & $0.6638$ & $0.6707$ & $\textbf{0.6807}$\\
& $\pm0.0008$ & $\pm0.0008$ & $\pm0.0006$ & $\pm0.0006$ & $\pm0.0006$ & $\pm0.0006$ & $\pm0.0006$\\
$G5_9$ & $0.6927$ & $0.6312$ & $0.5969$ & $0.6688$ & $0.6680$ & $0.6752$ & $\textbf{0.6866}$\\
& $\pm0.0008$ & $\pm0.0005$ & $\pm0.0006$ & $\pm0.0006$ & $\pm0.0006$ & $\pm0.0006$ & $\pm0.0006$\\

\bottomrule
\end{tabular}
\end{footnotesize}
\end{table*}

Before diving into experimental results, we revisit the
earlier example showing that {\sc MyerNet}/{\sc MinMax} style allocations may lead to significant revenue loss. Here we further elaborate on that example using specific numbers.
We assume that the bid profile is either $(1,0)$ or $(0.5,\epsilon)$, each with $50\%$ chance. That is, with $50\%$ chance, bidder $1$'s
value is $1$ and bidder $2$'s value is $0$, and with $50\%$
chance, bidder $1$'s value is $0.5$ and bidder $2$'s value is $\epsilon$, where $\epsilon$ is infinitesimal.
A near optimal mechanism works as follows: if bidder $2$'s value
is $0$, then bidder $1$ faces a take-it-or-leave-it offer of $1$,
and if bidder $2$'s value is $\epsilon$, then bidder $1$ faces a take-it-or-leave-it offer of $0.5$. The expected revenue is $0.75$.
Suppose our allocation follows the style of either {\sc MyerNet} or {\sc MinMax}. We use $p_0$ to represent bidder $1$'s critical price for winning when bidder $2$ bids $0$ and we use
$p_{\epsilon}$ to represent bidder $1$'s critical price for winning when bidder $2$ bids $\epsilon$.
Since we assume our allocation follows the style of either
{\sc MyerNet} or {\sc MinMax}, we must have $p_0\le p_{\epsilon}$.
For example, suppose the allocation follows the style of
{\sc MinMax}, then when bidder $2$'s bid increases from $0$
to $\epsilon$, bidder $1$'s priority value either stays the same
or drops and bidder $2$'s priority value either stays the same
or increases. In order for bidder $1$ to still beat bidder $2$,
bidder $1$'s minimum winning bid must either stay the same or increase. If $p_{\epsilon}\le 0.5$, then the maximum revenue extracted
from bidder $1$ is then at most $0.5p_0+0.5p_{\epsilon}\le 0.5$.
If $p_{\epsilon}> 0.5$, then the maximum revenue extracted
from bidder $1$ is then at most $0.5p_0\le 0.5$.
Therefore, the maximum revenue extracted from both bidders
is at most $0.5+\epsilon$. Earlier, we showed that the optimal revenue is at least $0.75$. That is, by adopting {\sc MyerNet} or {\sc MinMax}, for this example, we lose one third of the revenue.

In Table~\ref{tab:iclr}, we present experimental results on randomly generated grid distributions ($G5_0$ to $G5_9$) for $5$ bidders. We still
use the revenue gap (to the unattainable greedy upper bound) as the performance indicator.
Our method {\sc MP}'s revenue gap is $0.87\%$. The revenue gaps of {\sc MyerNet+} and {\sc MinMax+},
both are based on techniques proposed in our paper, are $3.3\%$
and $3.4\%$, respectively. For reference, we have also included the revenues of the manual baseline {\sc Myer.}, which is Myerson's greedy allocation
where virtual valuations are calculated based on the {\em marginal}
distributions (essentially ignoring correlation altogether). Both {\sc MyerNet+} and {\sc MinMax+} do not
outperform {\sc Myer.}
Lastly, {\sc MyerNet} and {\sc MinMax} have revenue gaps of $19\%$ and $18\%$.

\section{Multi-Unit Auctions with Unit Demand}
\label{sec:multi}

We extend our techniques to multi-unit auctions with unit demand.
We use $m$ to denote the number of items, where $m<n$.
Myerson's greedy allocation is extended by assigning
one item to each bidder with the $m$ highest marginal profit.

In the single-item setting, our neural network's output dimension is $n+1$. If the $i$-th ($i=1,2,\ldots, n$) output coordinate is the highest, then bidder $i$ wins. If the $(n+1)$-th coordinate is the highest, then the item is not allocated.
For the multi-unit setting, the network's output dimension remains $n+1$. Bidder $i$ wins if and only if the $i$-th output coordinate
is among the $m$ highest output coordinates and the $i$-th output coordinate is higher than the $(n+1)$-th output coordinate.
This allows us to represent all possible numbers of items allocated ({\em i.e.}, from $0$ to $m$).

The extended monotonicity condition requires that any winner $i$, who is in
the original {\em winner set}, must remains in the winner set when $i$ increases her bid while
the other bids stay the same.
This more complex version requires us to enumerate all combinations of $i$, all original winner sets that contain $i$ and all new winner sets that exclude $i$. The monotonicity verification MIP for the single-item setting
can be trivially adapted, as any given winner set can be described by a set of inequalities.

In Table~\ref{tab:multi}, we present our experimental results.
The baseline is the $(m+1)$-th price auction with the optimal reserve price.
For each distribution, we run $10$ trials, and in every trial, we stop after $5,000$, $10,000$, $15,000$ and $20,000$ optimizer steps to
perform monotonicity verification. We record the best-performing verifiably monotone allocations.

\begin{table*}[h]
\centering
    \caption{Multi-unit auctions with unit demand with $2$ items (i.e., $m=2$)\\
    {\sc m+1} refers to the (m+1)-th price auction with the optimal reserve price,
    $\textsc{Gap}(\textsc{MP})=0.67\%$,
    $\textsc{Gap}(\textsc{m+1})=5.1\%$
    }
  \label{tab:multi}
    \begin{footnotesize}
  \begin{tabular}{llllllllll}
    \toprule
      & \multicolumn{3}{c}{$n=3$} & \multicolumn{3}{c}{$n=4$}
      & \multicolumn{3}{c}{$n=5$}
      \\
     \cmidrule(r){2-4}
     \cmidrule(r){5-7}
     \cmidrule(r){8-10}
      {\sc Distrib.} &  {\sc Greedy} & {\sc m+1} & {\sc MP} & {\sc Greedy} & {\sc m+1} & {\sc MP} & {\sc Greedy} & {\sc m+1} & {\sc MP}\\
    \midrule
$G5_0$ & $0.7629$ & $0.7251$ & $\textbf{0.7624}$ &$0.9387$ & $0.8923$ & $\textbf{0.9360}$ &$1.0856$ & $1.0428$ & $\textbf{1.0757}$ \\
& $\pm0.002$ & $\pm0.001$ & $\pm0.001$ &$\pm0.002$ & $\pm0.001$ & $\pm0.001$ &$\pm0.002$ & $\pm0.001$ & $\pm0.001$ \\
$G5_1$ & $0.7896$ & $0.7413$ & $\textbf{0.7901}$ &$0.9646$ & $0.9172$ & $\textbf{0.9608}$ &$1.0973$ & $1.0477$ & $\textbf{1.0844}$ \\
& $\pm0.002$ & $\pm0.001$ & $\pm0.001$ &$\pm0.002$ & $\pm0.001$ & $\pm0.001$ &$\pm0.002$ & $\pm0.001$ & $\pm0.001$ \\
$G5_2$ & $0.7311$ & $0.6900$ & $\textbf{0.7311}$ &$0.9380$ & $0.8932$ & $\textbf{0.9319}$ &$1.0868$ & $1.0394$ & $\textbf{1.0755}$ \\
& $\pm0.002$ & $\pm0.001$ & $\pm0.001$ &$\pm0.002$ & $\pm0.001$ & $\pm0.001$ &$\pm0.002$ & $\pm0.001$ & $\pm0.001$ \\
$G5_3$ & $0.7497$ & $0.7048$ & $\textbf{0.7500}$ &$0.9344$ & $0.8853$ & $\textbf{0.9293}$ &$1.0930$ & $1.0462$ & $\textbf{1.0799}$ \\
& $\pm0.002$ & $\pm0.001$ & $\pm0.001$ &$\pm0.002$ & $\pm0.001$ & $\pm0.001$ &$\pm0.002$ & $\pm0.001$ & $\pm0.001$ \\
$G5_4$ & $0.7724$ & $0.7280$ & $\textbf{0.7733}$ &$0.9457$ & $0.8959$ & $\textbf{0.9397}$ &$1.0900$ & $1.0456$ & $\textbf{1.0737}$ \\
& $\pm0.002$ & $\pm0.001$ & $\pm0.001$ &$\pm0.002$ & $\pm0.001$ & $\pm0.001$ &$\pm0.002$ & $\pm0.001$ & $\pm0.001$ \\
$G5_5$ & $0.7511$ & $0.6956$ & $\textbf{0.7473}$ &$0.9329$ & $0.8841$ & $\textbf{0.9266}$ &$1.0938$ & $1.0463$ & $\textbf{1.0748}$ \\
& $\pm0.002$ & $\pm0.001$ & $\pm0.001$ &$\pm0.002$ & $\pm0.001$ & $\pm0.001$ &$\pm0.002$ & $\pm0.001$ & $\pm0.001$ \\
$G5_6$ & $0.7400$ & $0.6911$ & $\textbf{0.7401}$ &$0.9507$ & $0.8964$ & $\textbf{0.9449}$ &$1.0965$ & $1.0474$ & $\textbf{1.0852}$ \\
& $\pm0.002$ & $\pm0.001$ & $\pm0.001$ &$\pm0.002$ & $\pm0.001$ & $\pm0.001$ &$\pm0.002$ & $\pm0.001$ & $\pm0.001$ \\
$G5_7$ & $0.8122$ & $0.7734$ & $\textbf{0.8096}$ &$0.9663$ & $0.9201$ & $\textbf{0.9643}$ &$1.0945$ & $1.0498$ & $\textbf{1.0825}$ \\
& $\pm0.002$ & $\pm0.001$ & $\pm0.001$ &$\pm0.002$ & $\pm0.001$ & $\pm0.001$ &$\pm0.002$ & $\pm0.001$ & $\pm0.001$ \\
$G5_8$ & $0.7531$ & $0.6924$ & $\textbf{0.7525}$ &$0.9390$ & $0.8936$ & $\textbf{0.9349}$ &$1.0925$ & $1.0462$ & $\textbf{1.0785}$ \\
& $\pm0.002$ & $\pm0.001$ & $\pm0.001$ &$\pm0.002$ & $\pm0.001$ & $\pm0.001$ &$\pm0.002$ & $\pm0.001$ & $\pm0.001$ \\
$G5_9$ & $0.7949$ & $0.7329$ & $\textbf{0.7934}$ &$0.9682$ & $0.9130$ & $\textbf{0.9604}$ &$1.1016$ & $1.0523$ & $\textbf{1.0894}$ \\
& $\pm0.002$ & $\pm0.001$ & $\pm0.001$ &$\pm0.002$ & $\pm0.001$ & $\pm0.001$ &$\pm0.002$ & $\pm0.001$ & $\pm0.001$ \\
\bottomrule
  \end{tabular}
    \end{footnotesize}
\end{table*}

\section{Continuous Distributions}
\label{sec:continuous}

We present a few additional experiments on continuous distributions in Table~\ref{tab:iclrcontinuous}.
We experimented with three continuous distributions for $2$ bidders.
We use $f(x,y)$ to denote the joint probability density function. (We need to apply a normalisation factor $\alpha$ to ensure that $\int_{0}^1\int_{0}^1 \alpha f(x,y) dx dy=1$. We ignore $\alpha$ for presentation purpose.)
Our method {\sc MP} is near optimal. {\em Its gap to greedy is only on average $0.02\%$, also noting that the greedy revenue is not attainable.}
On the other hand, {\sc 2nd} (second price auction with optimal reserve) also performs quite well for all three cases.

\begin{table*}[h!]
\centering
    \caption{Experiments on three continuous distributions\\
    $\textsc{Gap}(\textsc{MP})=0.02\%$,$\textsc{Gap}(\textsc{2nd})=0.5\%$}
\label{tab:iclrcontinuous}
\begin{footnotesize}
\begin{tabular}{ll|ll}
\toprule
{\sc Distrib.} &  {\sc Greedy} & {\sc 2nd} & {\sc MP} \\
\midrule

    $f(x,y)=2x+y$ & $0.5095\pm 0.001$ & $0.5079\pm 0.0008$ & $\textbf{0.5094}\pm 0.0008$ \\
\midrule

    $f(x,y)=2x-y+1$ & $0.4518\pm 0.001$ & $0.4469\pm 0.0008$ & $\textbf{0.4517}\pm 0.0008$ \\
\midrule
    $f(x,y)=sin(x)+cos(y)+2$ & $0.4216\pm 0.001$ & $0.4211\pm 0.0008$ & $\textbf{0.4215}\pm 0.0008$ \\
\bottomrule
\end{tabular}
\end{footnotesize}
\end{table*}

\section{Distributions Based on Real-World Datasets}
\label{sec:realdata}

In Table~\ref{tab:realdata}, we evaluate our mechanism using empirical bid distributions derived from three real-world datasets. These datasets come from diverse real-world market domains: (i) \textbf{Consumer (eBay)}, involving {\bf layperson bidding} for electronics; (ii) \textbf{Financial (Ethereum)}, involving {\bf algorithmic bidding} for block building rights; and (iii) \textbf{Electricity Market (PJM)}, involving {\bf professional strategic bids} from energy producers.

\begin{itemize}
    \item \textbf{Consumer (eBay):}
        This dataset consists of historical eBay auction data for second-hand Palm PDAs, a market typically characterized by high valuation correlation, representing a \textit{layperson bidding} environment. From a total of 194 auctions sourced via \verb|modelingonlineauctions.com|, we identify a frequently co-appearing bidder pair (IDs {\sc CHIMAM} and {\sc DIDOJAN}) who participated in the same 13 auctions. These 13 joint appearances provide 13 bid profile samples, using their respective maximum bids in each instance. The bids are normalized to the $[0, 1]$ interval (relative to a \$200 maximum) and scattered into a $100 \times 100$ grid distribution.

    \item \textbf{Financial (Ethereum):}
        This dataset represents high-frequency auctions for block-building rights, featuring intense competition between sophisticated \textit{algorithmic bidders}. On May 1st, 2026, a total of 7,117 auctions were recorded, with the top two winners securing 1,200 and 976 auctions, respectively (their public keys end with {\sc 48e681} and {\sc 8eab08}). Notably, these two bidders co-appeared in 5,711 auctions. While this environment is highly contested, the shared nature of the financial rewards leads to a setting where second-price mechanisms already perform quite well, which is also confirmed in our experiments. We collect joint bid profiles from these two frequent winners, using their maximum bids in co-appearing auctions as samples. Bids exceeding 0.05 ETH are discarded. The remaining samples are normalized to $[0,1]$ and scattered into the $100\times 100$ grid distribution. Data is sourced from \verb|relayscan.io|.

    \item \textbf{Electricity Market (PJM):}
        Hourly wholesale auctions managed by the PJM Interconnection (a Regional Transmission Organization coordinating the largest competitive electricity market in the US) between Dec 1st 2025 and Jan 1st 2026. These auctions determine the \textit{generation rights of MWh for specific delivery hours}, featuring strategic bids from professionals on behalf of power producers. We extract joint bid profiles from the most frequently co-appearing bidder pair, filtering out ``static'' bidders who submit constant bids. The identified bidder pair (anonymized ID ending with {\sc yMjEwNA} and {\sc AyMjEw}) co-appear in 744 auctions, one bids between \$25.64 to \$72.45 and the other bids between \$77.75 to \$495.17.
We take the bidders' maximum hourly supply offers as bid samples. We normalize the bids to $[0,1]$ (maximum bid is $\$500$) and scatter them into the $100 \times 100$ grid distribution. Data is retrieved from the PJM Data Miner 2 repository \verb|dataminer2.pjm.com|.

\end{itemize}

\begin{table*}[h!]
\centering
    \caption{Experiments on Distributions Based on Real-World Datasets:\\
    $\textsc{Gap}(\textsc{MP})=2.0\%$,$\textsc{Gap}(\textsc{2nd})=22\%$,$\textsc{Gap}(\textsc{Myer.})=30\%$}
\label{tab:realdata}
\begin{footnotesize}
\begin{tabular}{ll|lll}
\toprule
    {\sc Distrib.} &  {\sc Greedy} & {\sc 2nd} & {\sc Myer.} & {\sc MP} \\
\midrule

    Consumer (eBay) & $0.5445\pm 0.0006$ & $0.4211\pm 0.0001$ & $0.3568\pm 0.0002$ & $\textbf{0.5373}\pm 0.0006$ \\
\midrule

    Financial (Ethereum) & $0.3040 \pm 0.0007$ & $0.2805\pm 0.0007$ & $0.2598\pm 0.0008$ & $\textbf{0.3022}\pm 0.0007$ \\
\midrule
    Electric Market (PJM) & $0.2508\pm 0.0005$ & $0.1578\pm 0.0001$ & $0.1575\pm 0.0001$ & $\textbf{0.2375}\pm 0.0004$ \\
\bottomrule
\end{tabular}
\end{footnotesize}
\end{table*}

The empirical results in Table~\ref{tab:realdata} demonstrate that our technique significantly outperforms manual baselines. The following analysis explains why these baselines fail.

\begin{itemize}
    \item Failure of the second-price auction with the optimal reserve ({\sc 2nd}): For eBay and PJM, there is a substantial distance between the highest and second-highest bids.
        Because the second bid is frequently much lower than the first (with gap of $35\%$ and $68\%$), the second price auction essentially functions as a single-bidder auction with a reserve, failing to extract the full value from the winning bidder.

    \item Failure of Myerson's technique based on marginal distributions ({\sc Myer.}): For eBay, there is a strong negative correlation (Pearson correlation coefficient $-0.87$) between the two bidders,
        while for PJM, there is a strong positive correlation (coefficient $0.85$) between the two bidders.
        {\sc Myer.} fails to adapt to the correlation patterns.
\end{itemize}

In contrast, our technique achieves near-optimal revenue for all three datasets.

\newpage
\section{Complete Tables Including Standard Errors}
\label{sec:fulltables}

Lastly, in Table~\ref{tab:n2-1} ($n=2$), \ref{tab:n3-1} ($n=3$), \ref{tab:n4-1} ($n=4$), \ref{tab:n5-1} ($n=5$), we present the complete experimental results including standard errors.

\begin{table*}[h!]
\centering
\caption{$2$ bidders (with standard errors)\\
$\textsc{Gap}(\textsc{MP})=0.26\%$,$\textsc{Gap}(\textsc{RNet})=2.7\%$,$\textsc{Gap}(\textsc{2nd})=4.2\%$,$\textsc{Gap}(\textsc{Myer})=4.3\%$}
\label{tab:n2-1}
\begin{footnotesize}
\begin{tabular}{lll|llllll}
\toprule
{\sc Distrib.} &  {\sc Greedy} & {\sc AMD} & {\sc Myer.} & {\sc 2nd} & {\sc RNet} & {\sc MP} & {\sc VV} & {\sc IVV} \\
\midrule

$G5_0$ & $0.4353$ & $0.4368$ & $0.4191$ & $0.4197$ & $0.4307$ & $\textbf{0.4368}$ & $0.4355$ & $0.4329$\\
& $\pm0.001$ & & $\pm0.0008$ & $\pm0.0009$ & $\pm0.0009$ & $\pm0.0009$ & $\pm0.0009$ & $\pm0.0009$\\
$G5_1$ & $0.4047$ & $0.4041$ & $0.3939$ & $0.3929$ & $0.4021$ & $\textbf{0.4037}$ & $0.4032$ & $0.4032$\\
& $\pm0.001$ & & $\pm0.0007$ & $\pm0.0006$ & $\pm0.0007$ & $\pm0.0007$ & $\pm0.0007$ & $\pm0.0007$\\
$G5_2$ & $0.3979$ & $0.3967$ & $0.3771$ & $0.3823$ & $0.3869$ & $\textbf{0.3959}$ & $0.3946$ & $0.3955$\\
& $\pm0.001$ & & $\pm0.0009$ & $\pm0.001$ & $\pm0.0009$ & $\pm0.001$ & $\pm0.001$ & $\pm0.001$\\
$G5_3$ & $0.4629$ & $0.4625$ & $0.4533$ & $0.4546$ & $0.4522$ & $\textbf{0.4628}^*$ & $0.4612$ & $0.4616$\\
& $\pm0.001$ & & $\pm0.0006$ & $\pm0.0006$ & $\pm0.0007$ & $\pm0.0006$ & $\pm0.0007$ & $\pm0.0007$\\
$G5_4$ & $0.4681$ & $0.4665$ & $0.4476$ & $0.4472$ & $0.4639$ & $\textbf{0.4674}^*$ & $0.4673$ & $0.4671$\\
& $\pm0.001$ & & $\pm0.0009$ & $\pm0.0009$ & $\pm0.0008$ & $\pm0.0008$ & $\pm0.0008$ & $\pm0.0008$\\
$G5_5$ & $0.4204$ & $0.4187$ & $0.4075$ & $0.4057$ & $0.4062$ & $\textbf{0.4151}$ & $0.4148$ & $0.4150$\\
& $\pm0.001$ & & $\pm0.0008$ & $\pm0.0008$ & $\pm0.0009$ & $\pm0.0009$ & $\pm0.001$ & $\pm0.001$\\
$G5_6$ & $0.3905$ & $0.3912$ & $0.3761$ & $0.3741$ & $0.3798$ & $\textbf{0.3908}$ & $0.3893$ & $0.3886$\\
& $\pm0.001$ & & $\pm0.0007$ & $\pm0.0008$ & $\pm0.0008$ & $\pm0.0007$ & $\pm0.0008$ & $\pm0.0008$\\
$G5_7$ & $0.4865$ & $0.4854$ & $0.4494$ & $0.4544$ & $0.4771$ & $\textbf{0.4850}$ & $0.4828$ & $0.4840$\\
& $\pm0.001$ & & $\pm0.0006$ & $\pm0.0008$ & $\pm0.0007$ & $\pm0.0008$ & $\pm0.0008$ & $\pm0.0008$\\
$G5_8$ & $0.4298$ & $0.4273$ & $0.3988$ & $0.3981$ & $0.4077$ & $\textbf{0.4291}^*$ & $0.4281$ & $0.4279$\\
& $\pm0.001$ & & $\pm0.001$ & $\pm0.0009$ & $\pm0.0009$ & $\pm0.0008$ & $\pm0.0009$ & $\pm0.0009$\\
$G5_9$ & $0.4531$ & $0.4512$ & $0.4306$ & $0.4302$ & $0.4379$ & $\textbf{0.4498}$ & $0.4482$ & $0.4497$\\
& $\pm0.001$ & & $\pm0.0009$ & $\pm0.0009$ & $\pm0.0009$ & $\pm0.0009$ & $\pm0.001$ & $\pm0.0009$\\
{\sc saturn} & $0.4533$ & $0.4545$ & $0.4210$ & $0.4226$ & $0.4372$ & $0.4510$ & $0.4512$ & $\textbf{0.4515}$\\
& $\pm0.001$ & & $\pm0.0008$ & $\pm0.0008$ & $\pm0.0008$ & $\pm0.0008$ & $\pm0.0008$ & $\pm0.0008$\\
{\sc jupiter} & $0.4427$ & $0.4404$ & $0.4238$ & $0.4219$ & $0.4316$ & $\textbf{0.4418}^*$ & $0.4409$ & $0.4410$\\
& $\pm0.001$ & & $\pm0.0008$ & $\pm0.0009$ & $\pm0.0008$ & $\pm0.0009$ & $\pm0.0009$ & $\pm0.0009$\\
{\sc mars} & $0.4375$ & $0.4377$ & $0.4177$ & $0.4167$ & $0.4194$ & $\textbf{0.4360}$ & $0.4333$ & $0.4359$\\
& $\pm0.001$ & & $\pm0.0008$ & $\pm0.0008$ & $\pm0.0008$ & $\pm0.0009$ & $\pm0.0009$ & $\pm0.0009$\\
{\sc sol} & $0.4429$ & $0.4414$ & $0.4376$ & $0.4364$ & $0.4329$ & $\textbf{0.4427}^*$ & $0.4404$ & $0.4419$\\
& $\pm0.001$ & & $\pm0.0008$ & $\pm0.0008$ & $\pm0.0009$ & $\pm0.0008$ & $\pm0.0008$ & $\pm0.0008$\\
{\sc venus} & $0.4381$ & $0.4381$ & $0.4175$ & $0.4171$ & $0.4233$ & $\textbf{0.4365}$ & $0.4319$ & $0.4358$\\
& $\pm0.001$ & & $\pm0.0008$ & $\pm0.0008$ & $\pm0.0008$ & $\pm0.0008$ & $\pm0.0009$ & $\pm0.0008$\\
{\sc mercury} & $0.4273$ & $0.4274$ & $0.4179$ & $0.4176$ & $0.4150$ & $\textbf{0.4281}^*$ & $0.4271$ & $0.4279$\\
& $\pm0.001$ & & $\pm0.0008$ & $\pm0.0008$ & $\pm0.0008$ & $\pm0.0008$ & $\pm0.0008$ & $\pm0.0008$\\
{\sc luna} & $0.4331$ & $0.4335$ & $0.4173$ & $0.4172$ & $0.4212$ & $\textbf{0.4322}$ & $0.4315$ & $0.4318$\\
& $\pm0.001$ & & $\pm0.0008$ & $\pm0.0008$ & $\pm0.0008$ & $\pm0.0008$ & $\pm0.0009$ & $\pm0.0009$\\

\bottomrule
\end{tabular}
\end{footnotesize}
\end{table*}

\begin{table*}[h!]
\centering
\caption{$3$ bidders (with standard errors)\\
    $\textsc{Gap}(\textsc{MP})=0.36\%$,
    $\textsc{Gap}(\textsc{RNet})=3.6\%$,
    $\textsc{Gap}(\textsc{2nd})=3.9\%$,
    $\textsc{Gap}(\textsc{Myer})=3.8\%$
    }
\label{tab:n3-1}
\begin{footnotesize}
\begin{tabular}{ll|llll}
\toprule
{\sc Distrib.} &  {\sc Greedy} & {\sc Myer.} & {\sc 2nd} & {\sc RNet} & {\sc MP} \\
\midrule

$G5_0$ & $0.5512$$\pm0.001$ & $0.5346$$\pm0.0007$ & $0.5338$$\pm0.0008$ & $0.5346$$\pm0.0007$ & $\textbf{0.5517}$$\pm0.0007$\\
$G5_1$ & $0.5647$$\pm0.0009$ & $0.5467$$\pm0.0007$ & $0.5475$$\pm0.0007$ & $0.5509$$\pm0.0007$ & $\textbf{0.5651}$$\pm0.0007$\\
$G5_2$ & $0.5316$$\pm0.001$ & $0.5137$$\pm0.0007$ & $0.5118$$\pm0.0008$ & $0.5153$$\pm0.0008$ & $\textbf{0.5302}$$\pm0.0008$\\
$G5_3$ & $0.5476$$\pm0.001$ & $0.5283$$\pm0.0007$ & $0.5282$$\pm0.0007$ & $0.5290$$\pm0.0007$ & $\textbf{0.5450}$$\pm0.0007$\\
$G5_4$ & $0.5485$$\pm0.0009$ & $0.5297$$\pm0.0007$ & $0.5306$$\pm0.0007$ & $0.5310$$\pm0.0007$ & $\textbf{0.5471}$$\pm0.0007$\\
$G5_5$ & $0.5477$$\pm0.001$ & $0.5237$$\pm0.0007$ & $0.5231$$\pm0.0007$ & $0.5255$$\pm0.0007$ & $\textbf{0.5443}$$\pm0.0008$\\
$G5_6$ & $0.5371$$\pm0.001$ & $0.5155$$\pm0.0008$ & $0.5138$$\pm0.0008$ & $0.5158$$\pm0.0007$ & $\textbf{0.5332}$$\pm0.0007$\\
$G5_7$ & $0.5723$$\pm0.0009$ & $0.5527$$\pm0.0007$ & $0.5528$$\pm0.0007$ & $0.5555$$\pm0.0006$ & $\textbf{0.5710}$$\pm0.0007$\\
$G5_8$ & $0.5473$$\pm0.001$ & $0.5186$$\pm0.0008$ & $0.5193$$\pm0.0007$ & $0.5185$$\pm0.0007$ & $\textbf{0.5429}$$\pm0.0008$\\
$G5_9$ & $0.5663$$\pm0.001$ & $0.5409$$\pm0.0007$ & $0.5410$$\pm0.0008$ & $0.5396$$\pm0.0007$ & $\textbf{0.5637}$$\pm0.0008$\\

\bottomrule
\end{tabular}
\end{footnotesize}
\end{table*}

\begin{table*}[h!]
\centering
\caption{$4$ bidders (with standard errors)\\
$\textsc{Gap}(\textsc{MP})=0.78\%$,
$\textsc{Gap}(\textsc{RNet})=3.5\%$,
$\textsc{Gap}(\textsc{2nd})=3.2\%$,
$\textsc{Gap}(\textsc{Myer})=3.2\%$}

\label{tab:n4-1}
\begin{footnotesize}
\begin{tabular}{ll|llll}
\toprule
{\sc Distrib.} &  {\sc Greedy} & {\sc Myer.} & {\sc 2nd} & {\sc RNet} & {\sc MP} \\
\midrule

$G5_0$ & $0.6270$$\pm0.0009$ & $0.6086$$\pm0.0007$ & $0.6082$$\pm0.0007$ & $0.6072$$\pm0.0006$ & $\textbf{0.6242}$$\pm0.0007$\\
$G5_1$ & $0.6417$$\pm0.0009$ & $0.6215$$\pm0.0006$ & $0.6205$$\pm0.0007$ & $0.6192$$\pm0.0006$ & $\textbf{0.6379}$$\pm0.0007$\\
$G5_2$ & $0.6250$$\pm0.0009$ & $0.6096$$\pm0.0007$ & $0.6101$$\pm0.0007$ & $0.6072$$\pm0.0006$ & $\textbf{0.6205}$$\pm0.0007$\\
$G5_3$ & $0.6266$$\pm0.0009$ & $0.6054$$\pm0.0006$ & $0.6055$$\pm0.0006$ & $0.6034$$\pm0.0006$ & $\textbf{0.6220}$$\pm0.0007$\\
$G5_4$ & $0.6296$$\pm0.0009$ & $0.6084$$\pm0.0006$ & $0.6077$$\pm0.0007$ & $0.6076$$\pm0.0006$ & $\textbf{0.6233}$$\pm0.0007$\\
$G5_5$ & $0.6213$$\pm0.0009$ & $0.6029$$\pm0.0007$ & $0.6028$$\pm0.0007$ & $0.6020$$\pm0.0006$ & $\textbf{0.6158}$$\pm0.0007$\\
$G5_6$ & $0.6347$$\pm0.0009$ & $0.6099$$\pm0.0007$ & $0.6085$$\pm0.0007$ & $0.6051$$\pm0.0006$ & $\textbf{0.6287}$$\pm0.0007$\\
$G5_7$ & $0.6427$$\pm0.0009$ & $0.6210$$\pm0.0006$ & $0.6234$$\pm0.0006$ & $0.6193$$\pm0.0006$ & $\textbf{0.6369}$$\pm0.0006$\\
$G5_8$ & $0.6286$$\pm0.0009$ & $0.6100$$\pm0.0007$ & $0.6091$$\pm0.0007$ & $0.6085$$\pm0.0007$ & $\textbf{0.6241}$$\pm0.0007$\\
$G5_9$ & $0.6411$$\pm0.0009$ & $0.6199$$\pm0.0006$ & $0.6201$$\pm0.0006$ & $0.6190$$\pm0.0006$ & $\textbf{0.6357}$$\pm0.0007$\\

\bottomrule
\end{tabular}
\end{footnotesize}
\end{table*}

\begin{table*}[h!]
\centering
\caption{$5$ bidders (with standard errors)\\
    $\textsc{Gap}(\textsc{MP})=0.87\%$,
    $\textsc{Gap}(\textsc{RNet})=2.8\%$,
    $\textsc{Gap}(\textsc{2nd})=2.4\%$,
    $\textsc{Gap}(\textsc{Myer})=2.4\%$}
\label{tab:n5-1}
\begin{footnotesize}
\begin{tabular}{ll|llll}
\toprule
{\sc Distrib.} &  {\sc Greedy} & {\sc Myer.} & {\sc 2nd} & {\sc RNet} & {\sc MP} \\
\midrule

$G5_0$ & $0.6862$$\pm0.0008$ & $0.6695$$\pm0.0006$ & $0.6700$$\pm0.0006$ & $0.6670$$\pm0.0006$ & $\textbf{0.6799}$$\pm0.0006$\\
$G5_1$ & $0.6901$$\pm0.0008$ & $0.6737$$\pm0.0006$ & $0.6738$$\pm0.0006$ & $0.6713$$\pm0.0005$ & $\textbf{0.6846}$$\pm0.0006$\\
$G5_2$ & $0.6838$$\pm0.0008$ & $0.6709$$\pm0.0006$ & $0.6691$$\pm0.0006$ & $0.6681$$\pm0.0005$ & $\textbf{0.6783}$$\pm0.0006$\\
$G5_3$ & $0.6894$$\pm0.0008$ & $0.6724$$\pm0.0006$ & $0.6721$$\pm0.0006$ & $0.6684$$\pm0.0005$ & $\textbf{0.6838}$$\pm0.0006$\\
$G5_4$ & $0.6853$$\pm0.0008$ & $0.6692$$\pm0.0006$ & $0.6682$$\pm0.0006$ & $0.6664$$\pm0.0006$ & $\textbf{0.6787}$$\pm0.0006$\\
$G5_5$ & $0.6850$$\pm0.0008$ & $0.6717$$\pm0.0006$ & $0.6720$$\pm0.0006$ & $0.6685$$\pm0.0006$ & $\textbf{0.6809}$$\pm0.0006$\\
$G5_6$ & $0.6902$$\pm0.0008$ & $0.6710$$\pm0.0006$ & $0.6725$$\pm0.0006$ & $0.6686$$\pm0.0006$ & $\textbf{0.6839}$$\pm0.0006$\\
$G5_7$ & $0.6900$$\pm0.0008$ & $0.6719$$\pm0.0006$ & $0.6715$$\pm0.0006$ & $0.6682$$\pm0.0005$ & $\textbf{0.6824}$$\pm0.0006$\\
$G5_8$ & $0.6868$$\pm0.0008$ & $0.6707$$\pm0.0006$ & $0.6715$$\pm0.0006$ & $0.6659$$\pm0.0006$ & $\textbf{0.6807}$$\pm0.0006$\\
$G5_9$ & $0.6927$$\pm0.0008$ & $0.6752$$\pm0.0006$ & $0.6768$$\pm0.0006$ & $0.6711$$\pm0.0005$ & $\textbf{0.6866}$$\pm0.0006$\\

\bottomrule
\end{tabular}
\end{footnotesize}
\end{table*}

\end{document}